\documentclass[onecolumn]{revtex4}
\usepackage{amsfonts, amsmath, bm, amssymb,graphicx}

\newcommand{\beq}{\begin{equation}}
\newcommand{\enq}{\end{equation}}

\newcommand\I{\mbox{i}}
\newcommand\D{\mbox{d}}

\newcommand{\e}{\eqref}

\newcommand{\p}{\partial}

\begin{document}

%%%% Article title to be placed here
\title{New conformal mapping for adaptive resolving of the complex singularities of Stokes wave
}

\author{%%%% Author details
 Pavel M. Lushnikov$^{1}$, Sergey A. Dyachenko$^{2}$ and Denis A. Silantyev$^{1}$}
\affiliation{$^{1}$ Department of Mathematics \& Statistics, University of New Mexico, Albuquerque, NM, USA\\
$^{2}$Institute for Computational and Experimental Research in Mathematics at Brown University, Providence, RI, USA
}

%\author{%%%% Author details
% Pavel M. Lushnikov}
%\affiliation{Department of Mathematics \& Statistics, University of New Mexico, Albuquerque, NM, USA}
%%\email[]{plushnik@math.unm.edu}
%
%\author{Sergey A. Dyachenko$^{*}$}
%\affiliation{Institute for Computational and Experimental Research in Mathematics at Brown University, Providence, RI, USA}
%\email[]{sdyachen@math.uiuc.edu}
%
%\author{Denis A. Silantyev}
%\affiliation{Department of Mathematics \& Statistics, University of New Mexico, Albuquerque, NM, USA}

%%%%%%%%% Insert author address here

%%%% Subject entries to be placed here %%%%
%\subject{fluid dynamics, water waves, conformal mapping}

%%%% Keyword entries to be placed here %%%%

%%%% Insert corresponding author and its email address}
%\corres{P. M. Lushnikov\\

%%%% Abstract text to be placed here %%%%%%%%%%%% We introduce an auxiliary conformal mapping that allows
\begin{abstract}
A new highly efficient method is developed for computation of traveling periodic waves (Stokes waves) on the free surface of deep water.
A convergence of numerical approximation  is  determined by the complex singularites above the free surface for the analytical continuation of the travelling wave  into the complex plane.   An auxiliary conformal mapping is introduced which moves  singularities away from the free surface thus dramatically speeding up numerical convergence by adapting the numerical grid for resolving  singularities while being consistent with the fluid dynamics.
The efficiency of that conformal mapping is demonstrated for Stokes wave approaching  the limiting Stokes wave (the wave of the greatest height) which significantly expands the family of numerically accessible solutions.
It allows to provide a detailed study of the oscillatory approach of these solutions to the limiting wave.
 Generalizations of the conformal  mapping to resolve multiple singularities are also introduced.
\end{abstract}
%%%%%%%%%%%%%%%%%%%%%%%%%%%

%%%%%%%%%% Insert the texts which can accomdate on firstpage in the tag "fmtext" %%%%%

%\begin{fmtext}
%Placeholder for the text in front page
%\jname{rspa}
%\Journal{Proc R Soc A\ }
%\end{fmtext}

%%%%%%%%%%%%%%% End of first page %%%%%%%%%%%%%%%%%%%%%

\maketitle

\section{Introduction}
\label{sec:introduction}

The potential flow of ideal fluid of infinite depth with free surface can be   efficiently described through the time-dependent conformal mapping
\begin{equation} \label{zwdef}
z(w,t)=x(w,t)+\I y(w,t)
\end{equation}
of the lower complex half-plane $\mathbb{C}^-:=\{w|v\le 0\}$ of the auxiliary complex variable
\begin{equation} \label{wdef}
w:=u+\I v, \quad -\infty<u<\infty,
\end{equation}
into the area $-\infty<x<\infty,$ $ y\le \eta(x,t)$ occupied by the fluid
\cite{Ovsyannikov1973,MeisonOrzagIzraelyJCompPhys1981,TanveerProcRoySoc1991,TanveerProcRoySoc1993,DKSZ1996,ZakharovDyachenkoVasilievEuropJMechB2002}, where  $y=\eta(x,t)$  is
 the coordinate of the free surface, $t$ is the time,
$x$ and $y$ are the horizontal and vertical physical coordinates,
respectively. Here the real line $v=0$ is mapped into the line
$y=\eta(x,t)$   representing the free surface of the fluid (see Fig. 1 of
Ref. \cite{DyachenkoLushnikovKorotkevichPartIStudApplMath2016} for
the schematic of the conformal mapping \e{zwdef}).  Refs.
\cite{DyachenkoLushnikovKorotkevichJETPLett2014,DyachenkoLushnikovKorotkevichPartIStudApplMath2016,LushnikovStokesParIIJFM2016}
used this conformal transformation extensively, both
analytically and numerically,  to reveal the structures of complex
singularities of Stokes wave which is  the fully  nonlinear periodic
gravity wave propagating with the constant velocity $c$
~\cite{Stokes1847,Stokes1880}.
 Nonlinearity of Stokes wave
increases with the increase of $H/\lambda,$ where  $H$ is the Stokes
wave height which is defined as the vertical distance from the crest
to the trough of Stokes wave.  We use
scaled units  at which without the loss of generality the
spatial period is $\lambda=2\pi$ and $c=1$ for the linear gravity waves
similar to Ref. \cite{DyachenkoLushnikovKorotkevichPartIStudApplMath2016}. In a Stokes wave $c>1$ and the limit  $H\to 0,
\ c\to 1$ corresponds to the  linear gravity wave.  The
Stokes wave of the greatest height  $H=H_{max}$ (also known as the
limiting Stokes wave) has an angle of $2\pi/3$ radians at the crest,
corresponding to a singularity $z \sim w^{2/3}$~\cite{Stokes1880sup}.
%the singularity in the form  of the sharp
%angle of $2\pi/3$ radians on the crest~\cite{Stokes1880sup}.
%
The non-limiting Stokes waves describe ocean swell and the slow time
evolution of the Stokes wave toward its limiting form is one of the
possible routes to wave-breaking and whitecapping in full wave
dynamics. Wave-breaking and whitecapping carry away significant part
of energy and momenta of gravity waves~\cite{ZKPR2007, ZKP2009}.
%
%while  a slow    approaching of Stokes wave over time to its
%limiting form during wave dynamics is one of the possible routes to
%wave breaking and whitecapping, which are responsible for
%significant part of energy dissipation for gravity
%waves~\cite{ZKPR2007, ZKP2009}.
%
Here slow approach means the time scale which is much larger than
the temporal period of the gravity wave of the same spatial period
as for the given Stokes wave.  Formation of  limiting Stokes wave is
also considered to be a probable final stage of evolution of a freak
(or rogue) waves in the ocean resulting in formation of approximate
limiting Stokes wave for a limited period of time with following
wave breaking and disintegration of the wave or whitecapping and
attenuation of the freak wave into wave of regular amplitude
\cite{ZakharovDyachenkoProkofievEuropJMechB2006,RaineyLonguet-HigginsOceanEng2006,DyachenkoNewell2016}.

Thus the approach of non-limiting Stokes wave to the limiting Stokes wave has both  significant theoretical and practical interests. It was studied in details
in Refs. \cite{DyachenkoLushnikovKorotkevichJETPLett2014,DyachenkoLushnikovKorotkevichPartIStudApplMath2016,LushnikovStokesParIIJFM2016} how the complex
singularity in $w$ plane approaches the real line (corresponds to the fluid's free surface)  from above during the transition from non-limiting Stokes wave
to the limiting Stokes. Describing such transition is a numerically challenging task because in a Stokes wave the distance $v_c$ between the lowest branch points
to the real line approaches zero, which implies slow decay of the Fourier coefficients:
\begin{equation} \label{expspectral}
\hat{\tilde{z}}_k \propto e^{-v_c |k|} \quad  \text{for} \quad k\gg 1,
\end{equation}
 where $k$ is the Fourier wavenumber. Here, similar to Refs. \cite{DKSZ1996,DyachenkoLushnikovKorotkevichJETPLett2014,DyachenkoLushnikovKorotkevichPartIStudApplMath2016} we separated $z(w,t)$ into a $2\pi$-periodic part $\tilde{z}(2\pi+w,t)=\tilde{z}(w,t)$ and a non-periodic part $w$ by introducing
\begin{equation} \label{zwdef3}
\tilde{z}(w,t)\equiv z(w,t)-w=\tilde{x}(w,t)+\I y(w,t) \quad \text{with} \quad \tilde{x}(w,t)\equiv x(w,t)-w
\end{equation}
such that $\tilde{x}(-\pi,t)=\tilde{x}(\pi,t)=0$ while $x(-\pi,t)=-\pi$ and $x(\pi,t)=\pi$.
Ref.  \cite{DyachenkoLushnikovKorotkevichPartIStudApplMath2016} used up to   $M=2^{27}\approx 134\times10^{6}$ Fourier modes for  $\hat{\tilde{z}}_k$   on the uniform grid  which allowed to obtain the Stokes wave with $v_c = 5.93824419892803271779\ldots  \times10^{-7}$ (the maximal Fourier mode $k_{max}$ resolved in these simulations corresponds to $k_{max}=M$).

Conformal mappings can be used for improving efficiency of simulations  for the general periodic 1D system defined on the real line if such system allows  analytic continuation to the strip containing the real axis (see e.g. \cite{BoydChebyshevFourierSpectralMethodsBook2001} for review). Assume that $v_c$ is the vertical distance from the real line to the complex singularity closest to the real line. Thus  $v_c$ defines the thickness of the strip of analyticity in the direction where the singularity is nearest to the real line (Stokes wave is a special case because the thickness of strip is determined by the distance $v_c$ in   the upper complex half-plane $w\in \mathbb{C}^+$ while the thickness is infinite  below the real line). Then the FT for the system scales as in Eq. \e{expspectral}. The idea is to find a conformal transformation from $w$ to the new complex variable which makes the strip of analyticity thicker, i.e. to  push  all complex singularities of the system to the distance $\tilde v_c>v_c$ from the real line. Then FT in the new conformal variable  scales as $\propto e^{-\tilde v_c |k|} $  for $k\gg 1,$ i.e. decays faster than in  Eq.  \e{expspectral} speeding up numerical convergence.   A similar idea can be applied to the nonperiodic systems holomorphic in a closed ellipse around the segment of the real line (with foci corresponding to the two ends of that segment) with e.g.  rational spectral interpolants used instead of FT  \cite{TeeTrefethenSIAMJSCICOMPUT2006}.   In all such cases the spectral numerical methods including FT methods are highly efficient and typically having exponential convergence with the number of grid points $M$ used for the spectral collocation as exemplified  by Eq. \e{expspectral} if we use  $k=k_{max}=M$ for the estimate of the numerical error.   However exploiting such idea for the dynamics of the ideal fluid with free surface on infinite depth   has previously met with obstacles because the water waves dynamics require to work with function holomorphic in the entire lower complex half plane instead of the strip. In other words, only  singularities in the upper complex half-plane $w\in \mathbb{C}^+$ are allowed for the dynamics of the ideal fluid.
In this paper we overcome that obstacle by the proper choice of the conformal map.

We focus on problems periodic in $x$ variable (with period $2\pi$) in the reference frame moving with the constant velocity $c$. The
transformation \e{zwdef} becomes independent of time:
\begin{equation} \label{zwdef2}
z(w)=x(w)+\I y(w)
\end{equation}
which parametrically defines the Stokes wave as $y(x)$ in physical coordinates with $x(u)$ and $y(u)$ being horizontal and vertical physical coordinates, respectively.

% w and q plane figure
\begin{figure}
\includegraphics[width = 0.9\textwidth]{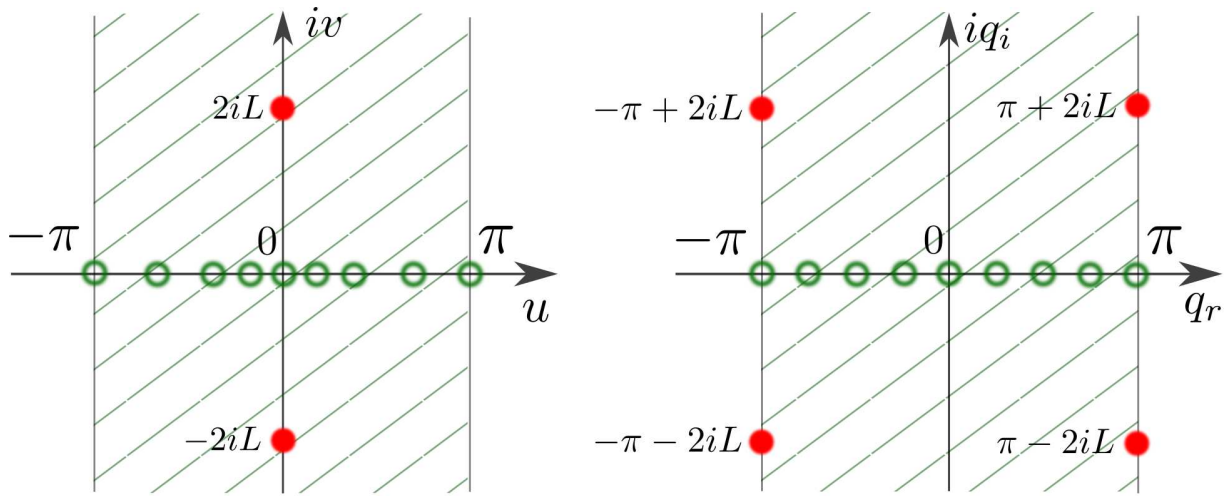}
\caption{A schematic of the conformal map ~\eqref{qnewdef} from the
strip $-\pi\le Re(w)<\pi$ of the complex variable $w\equiv u+\I v$
(left panel) into the strip  $-\pi\le Re(q)<\pi$ of the complex
variable $q\equiv q_r+\I q_i$ (right panel). The only exceptions are
the singularities of the conformal map at  $w=\pm2 \arctan(\I
L)+2\pi n=\pm2\I L+2\pi n +O(\I L^3), \ n=0,\pm1.,\pm2,\ldots $ (shown
by filled circles at left panel) which are mapped to the complex
infinity in $q$. In a similar way, the inverse conformal map from
$q$ to $w$  is singular at points   $q=\pm2 \arctan(\I/ L)+\pi(2
n+1)=\pm2\I L+\pi(2n+1) +O(\I L^3), \ n=0,\pm1.,\pm2,\ldots $ (shown by
filled circles at right panel) which are mapped to the complex
infinity in $w$.  Open circles schematically show that the uniform grid
(uniformly spaced points) at the real line $q=Re(q)$   is mapped
into the nonuniform grid at the real line $w=Re(w)$. The nonuniform
grid is denser near $w=0$ which allows to adaptively resolve the
complex singularity (branch point) of Stokes wave located at the
imaginary axis $w=\I v_c.$
 }
\label{fig:w-and-q}
\end{figure}

% Transform u(q) for L=0.0186088
\begin{figure}
\includegraphics[width = 0.495\textwidth]{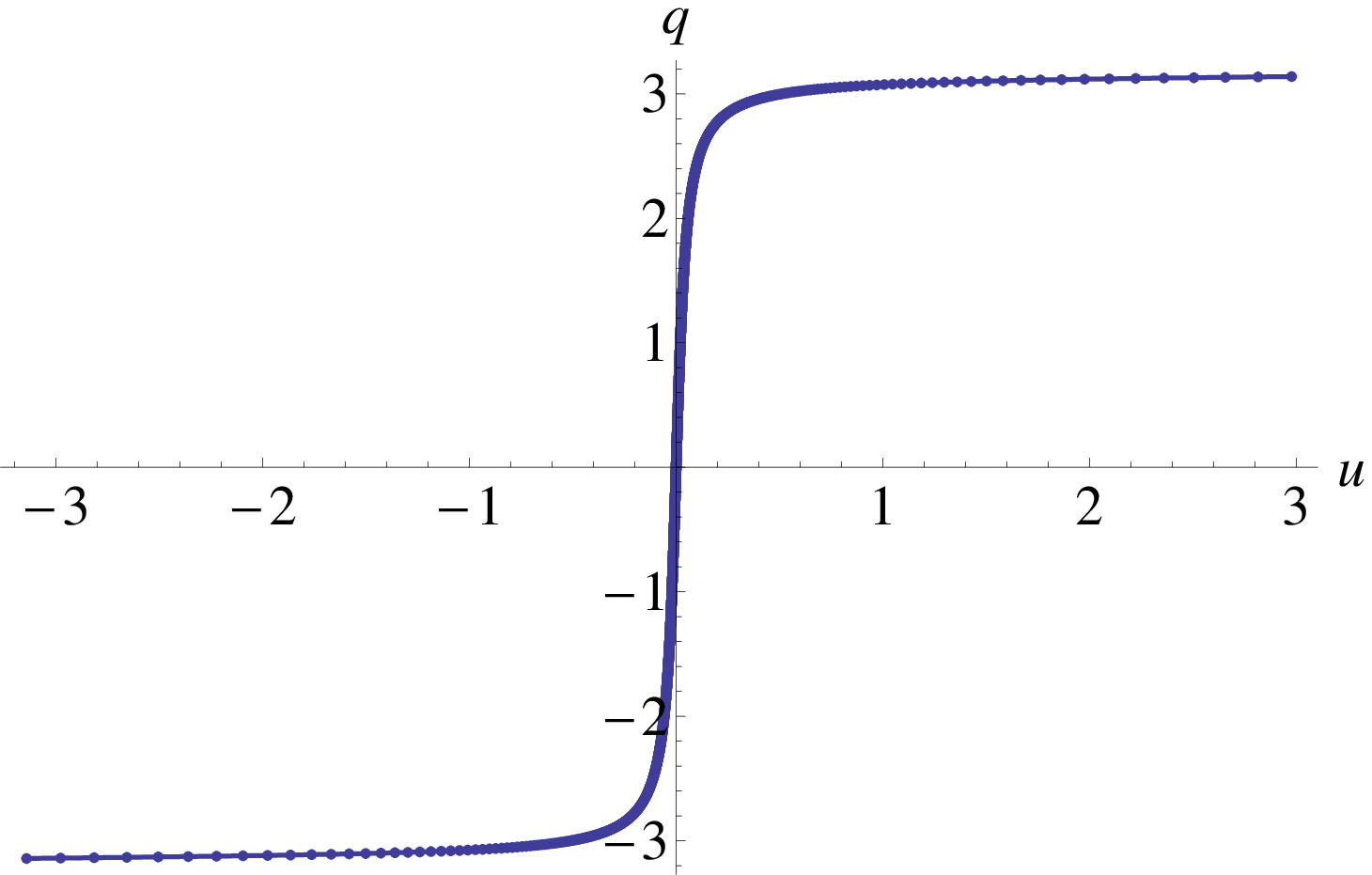}
\includegraphics[width = 0.495\textwidth]{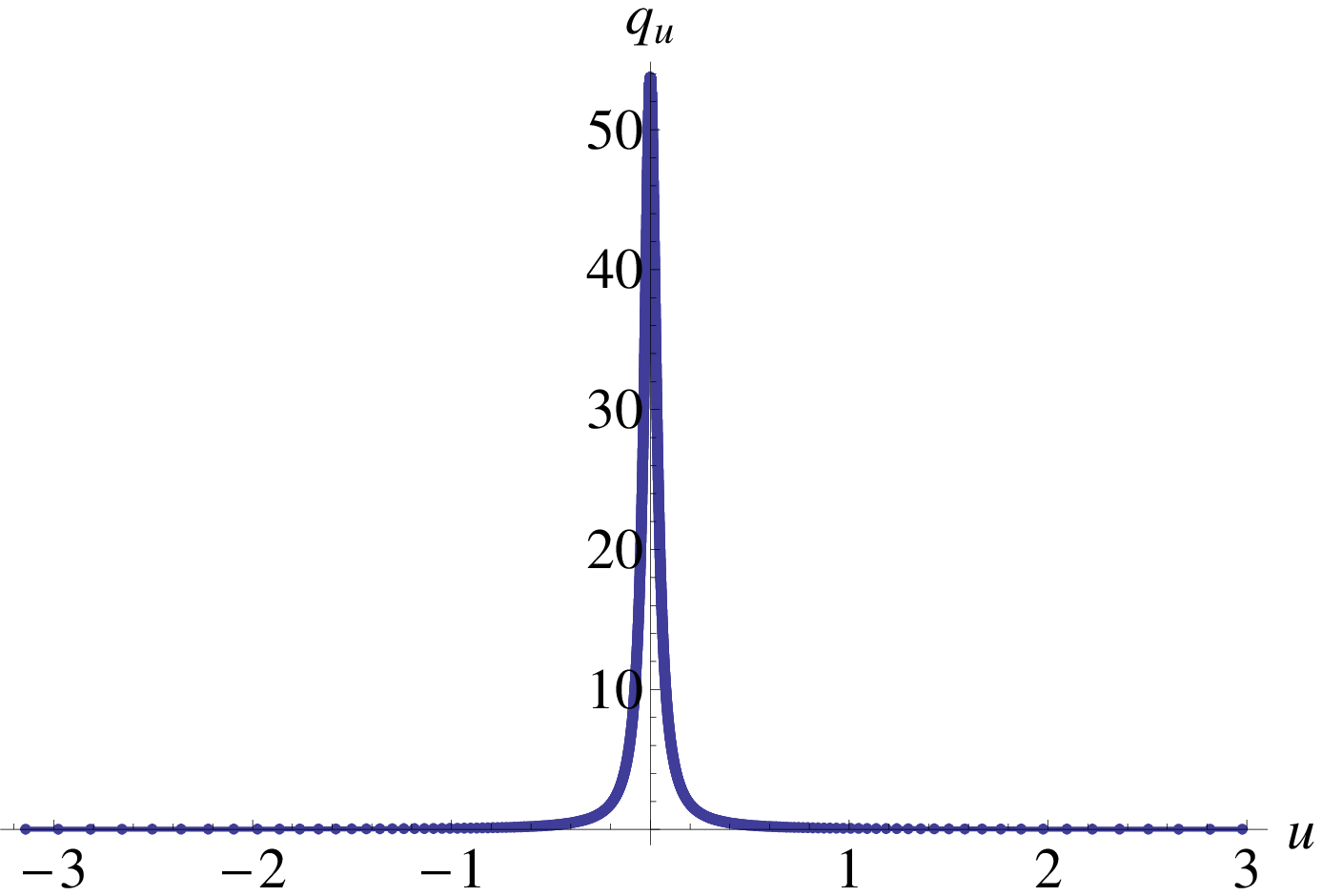}
\caption{The conformal  map  $q(u)$ given by equation \e{qnewdef}  (left panel) and the Jacobian $q_u(u)$ given by equation \e{dqdugeneral} (right panel) with $L=0.0186...$. The dots represent $M_q=1024$ points which are uniformly spaced at the interval $-\pi\le q<\pi$  the variable $q$ while being strongly concentrated near $u=0$ in the variable $u$. }
\label{fig:transform}
\end{figure}

The main results of this paper is that we found  a new conformal map
\begin{equation} \label{qnewdef}
q=2\,\text{arctan}{\left [ \frac{1}{L}\tan{\frac{w}{2}}  \right]},
\end{equation}
%
%shown in  Fig. \ref{fig:transform}
which we demonstrate to be consistent  with the  fluid dynamics. We show below the extreme  efficiency of equation \e{qnewdef} for simulations of Stokes wave. Here   $q$ is the new complex coordinate and $L$ is the arbitrary positive constant which we adjust to optimize performance of simulations. For the general time-dependent problems $L$ would become time-dependent to account for the vertical motion of nearest singularities as well as both $w$ and $q$ can be translated in the horizontal direction with time to account for the horizontal motion of singularities. However such generalization to time-dependent problems is beyond the scope of this paper.

Figs. \ref{fig:w-and-q} and \ref{fig:transform} show schematically that the new conformal map   \e{qnewdef} zooms at the real line into the
neighborhood of $w=0.$ Among all points on the real line $w=Re(w),$ that point $w=0$ is the closest  to the lowest singularity (branch cut) of the strongly nonlinear Stokes wave which is located at $w=\I v_c, \ v_c\ll 1$.  Then the uniform grid in the new variable $q$ corresponds to the highly nonuniform grid in the physical coordinates with the grid points concentrating at the neighborhood of the singularity
as seen in Fig. \ref{fig:transform}. The substitution of  $w=\I v_c$ \  into  equation   \e{qnewdef} immediately reveals that the lowest singularity of Stokes wave  is located at $q=\I v_c/L+O(\I v_c^3/L^3)\simeq\I v_c/L$ in $q$ plane.  It means that  the free parameter $L$ of the transformation \e{qnewdef} allows to change the position of the singularity in the complex $q$ plane. Here it is assumed that $v_c/L\ll 1$. Then FT in $q$ variable decay  as
\begin{equation} \label{qFT}
\hat{\tilde{z}}_k \propto e^{-(v_c/L) |k|} \quad  \text{for} \quad k\gg 1
\end{equation}
which is much faster than \e{expspectral}  for $L\ \ll 1.$ It makes the new conformal map  \e{qnewdef} highly efficient. Equation  \e{qnewdef}  has its own singularities at    $q=\pm2 \arctan(\I/ L)+2n\pi=\pm2\I L+(2n+1)\pi +O(\I L^3), \ n=0,\pm1.,\pm2,\ldots $ which approach the real line with the decrease of $L $ as schematically shown in Fig. \ref{fig:w-and-q}.  Balancing the contribution of singularities of Stokes wave and          $q(w)$ (i.e. setting them to have the same distance to the real axis in $q$ plane)  one obtains the optimal value
\begin{equation} \label{Loptimal}
L_{optimal}\simeq \left (\frac{v_c}{2} \right )^{1/2}
\end{equation}
which ensures the fastest possible convergence of Fourier modes as
\begin{equation} \label{vc1p2}
\hat{\tilde{z}}_k \propto e^{-(2v_c)^{1/2} |k|} \quad \text{for} \quad k\gg 1.
\end{equation}

E.g., the simulation of Ref.  \cite{DyachenkoLushnikovKorotkevichPartIStudApplMath2016} with $M=2^{27}\simeq1.3\cdot 10^8$ and  $v_c = 5.93824419892803271779\ldots  \times10^{-7}$ required running 64 cores computer cluster  for $\sim 3$ months. In contrast, the simulations  described in Section \ref{sec:results}     (they use  the new conformal map  \e{qnewdef}) allowed to achieve the same precision for the numerical grid with $M_{q}\simeq4.2\cdot10^4$  Fourier modes which takes a few minutes on the desktop computer. Respectively, by increasing $M_q$ (according to equation \e{vc1p2}, one has to choose  $M_q\sim M^{1/2}$ to reach the same precision as on  the uniform grid)  we were able to study Stokes waves with significantly smaller values of $v_c$ (down to $\sim10^{-11}$) than in Ref.  \cite{DyachenkoLushnikovKorotkevichPartIStudApplMath2016}.

\begin{figure}
\includegraphics[width = 1\textwidth]{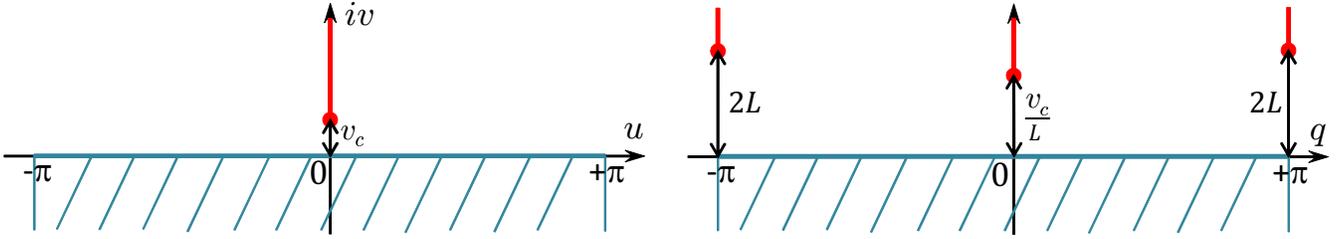}
\caption{Dots schematically show the singularity  at $w=\I v_c$ of
Stokes wave in the  variable $w=u+\I v$ (left panel) and the same
singularity at $q=\I v_c/L+O(\I v_c^3/L^3)\simeq\I v_c/L$ together with
the new singularities due to the inverse of the conformal map
\e{qnewdef}   at $q=\pm\pi+2 \I L+O(\I L^3)\simeq\pm\pi+2 \I L$ in the
new variable $q$ (right panel). The line extended upwards  to
$\I\infty$ on the left panel corresponds to branch cut of the Stokes
wave.  That branch cut is mapped by  \e{qnewdef}    into  three
branch cuts extending upwards  to $\I\infty$ from three dots on
right panel with  $\I\infty$ from $w$ plane mapped into both $q=\pm\pi+2 \I
L+O(\I L^3)$ by $2\pi$ periodicity.  See the text for more details on
that. Shaded areas on both panels correspond to the area occupied by
fluid.  } \label{fig:singularities}
\end{figure}

The new conformal map   \e{qnewdef} and its inverse provide the
mapping between  half-strips in $w$ and $q$ lower complex planes as
shown in Fig.   \ref{fig:singularities} by shaded areas. These areas
extend all way down in the complex planes and correspond to the area
occupied by fluid with the exception of the singularity points of
the conformal map as detailed in Section
\ref{sec:coordinatenonuniform}. These exceptional points  result in
the extra constant terms found in Section
\ref{sec:ProjectorsHilberttransformationq} to ensure the exact
solution of Euler equation through the conformal map.

% Singularities figure

The paper is organized as follows. In Section
\ref{sec:coordinatenonuniform} we discuss the properties of the
conformal map    \e{qnewdef} in complex plane and  corresponding
discretization. In Secton \ref{sec:StokesWave} we introduce the
closed equation for Stokes wave in new $q$ variable. In Section
\ref{sec:ProjectorsHilberttransformationq} we analyze how to work
with the projectors and Hilbert transformation in the new variable
$q$. In Section \ref{sec:stokeswavecomputation} we describe the
numerical algorithm used for obtaining the Stokes waves in the limit
of the complex singularity approaching the real line. Section \ref{sec:results}  demonstrates
the high efficiency of the new conformal map \e{qnewdef} and analyzes the results on the computed Stokes
waves. Section
\ref{sec:Generalizedconformalmap} provides a generalization of
the conformal map   \e{qnewdef} to adaptively resolve multiple
singularities. In Section \ref{sec:Conclusion} the main results of
the paper are discussed.

\section{New spatial coordinate for non-uniform grid}
\label{sec:coordinatenonuniform}

The conformal transformation  \e{qnewdef}  is  $2\pi$-periodic in $w$. Inverting equation \e{qnewdef} at the real line $u=w,$ we obtain that
\begin{equation} \label{unewdef}
u=2\,\text{arctan}{\left [ L\tan{\frac{q}{2}}  \right]}
\end{equation}
which is $2\pi$-periodic in $q$. Also the real line $w=u$ maps into the real line $q=Re(q)$. Recalling that we assume $2\pi$ periodicity of Eq. \e{zwdef3} for Stokes wave, we conclude that it is sufficient to consider the conformal transformations   \e{qnewdef}    and \e{unewdef} between  half-strip $-\pi     \le u\le \pi$, $-\infty<v\le 0$
and  $-\pi     \le Re(q)\le \pi$, $-\infty<Im(q)\le 0$ in $w$ and $q$, respectively.  Here  $2\pi$-periodicity is ensured by the limits $q\to \pm\pi$ for $u\to\pm\pi.$

If we assume that $|u|\ll L$ then equation \e{qnewdef} is reduced to
\begin{equation} \label{usmall1}
q=\frac{u}{L}
\end{equation}
which implies that taking numerical step $\Delta q\sim 1$ in $q$ space for $q$ near 0 is equivalent to taking the numerical step $\Delta u\sim L\ll1$ in $u$ space. It ensures that the uniform grid in $q$ space is highly concentrated near $u=0$ in $u$ space, with a "density" of grid points $q_u\sim1/L$ near $u=0$. It  allows to use much less grid points on the uniform grid in $q$ space in comparison with the uniform grid in $u$ space (the uniform grid was used previously in many simulations, see e.g. Refs.
\cite{ZakharovDyachenkoVasilievEuropJMechB2002,ZakharovDyachenkoProkofievEuropJMechB2006,DyachenkoLushnikovKorotkevichJETPLett2014,DyachenkoLushnikovKorotkevichPartIStudApplMath2016}) to archive the same precision of a numerical solution. To make these arguments  precise we use the Jacobian $q_u$ of the transformation \e{qnewdef} is given by
\begin{equation} \label{qjacobian}
q_u=\frac{1}{u_q}=\frac{1}{L\cos^2{\frac{u}{2}}\,
\left(1+\frac{1}{L^2}\tan^2{\frac{u}{2}}\right )}=\frac{\cos^2{\frac{q}{2}}\,
\left(1+L^2\tan^2{\frac{q}{2}}\right )}{L}=\frac{1+L^2+(1-L^2)\cos{q}}{2L}.
\end{equation}
Then for a general value of $q$, the steps $\Delta q$ and $\Delta u$ are related as
\begin{equation} \label{dqdugeneral}
\Delta q=q_u\Delta u+O(\Delta u^2)=\frac{1+L^2+(1-L^2)\cos{q}}{2L}\Delta u+O(\Delta u^2).
\end{equation}
Fig. \ref{fig:transform} shows  $q(u)$ and $q_u$ for $L=0.0186\ldots$ with dots representing $M_q=1024$ points of discrete grids both in $u$ and $q$ spaces separated by $\Delta u$ and $\Delta q$, respectively.

The branch point singularity of Stokes wave, located at $w=\I v_c$  in $w$ plane, corresponds to $q=\I q_c$ in $q$ plane in accordance with the Eq.   \e{qnewdef} where%
\begin{equation} \label{qcdef}
q_c=2\,\text{arctanh}{\left [ \frac{1}{L}\tanh{\frac{v_c}{2}}  \right]}
\end{equation}
which implies that
\begin{equation} \label{qcdefser}
q_c=\frac{v_c}{L}+O\left (\frac{v_c^3}{L^3}\right )\simeq \frac{v_c}{L} \quad \text{for} \quad v_c\ll L \quad
\end{equation}
as schematically shown in Fig. \ref{fig:singularities}.
Thus $q_c$ is located significantly higher in $q$ plane compared to $v_c$ in $w$ plane providing the quantitative explantation of much quicker decay of Fourier spectrum \e{qFT} in $q$ variable compared to Eq. \e{expspectral}. However, the asymptotic  \e{qFT} is valid provided $q=\I q_c$ is closer to the real axis than the other parts of the mapping of the Stokes wave branch cut $w\in[\I v_c, \I\infty)$ into $q$ plane.  In particular, a one part $w\in[\I v_c,2\I L+O(\I L^3))  $ of Stokes wave branch cut is mapped into $q\in[\I q_c,\I \infty)$ and another part  $w\in(2\I L+O(\I L^3),\I\infty)$ is mapped into two branch cuts $q\in(\pm\pi+\I \infty, \pm\pi+2\I L+O(\I L^3))$ by $2\pi$ periodicity in $q$ as sketched by vertical lines in Fig. \ref{fig:singularities}. Here the branch points $\pm\pi+2\I L+O(\I L^3)$ correspond to the  singularities of  the conformal map  \e{unewdef} in $q$ space.

These singularities are obtained from the Jacobian \e{qjacobian}
which is nonsingular for any value  $q\in \mathbb{C}$ but  reaches zero (i.e. the singularity of  $u_q$)  at
\begin{equation} \label{quzero}
q\:=q_{\pm}:=\pm 2\arctan{\frac{\I}{L}}+2n\pi, \quad n=0,\pm1,\pm 2,\ldots.
\end{equation}
For $L\ll 1$ Eq. \e{quzero} reduces in the strip $-\pi\le Re(q)\le \pi$ to
\begin{equation} \label{quzeroser}
q_{\pm}=\pm \pi\pm 2\I L +O(\I L^3) .
\end{equation}
The locations $q_{+}$ and $\I q_c$ of Stokes wave singularities in $q$ space are shown schematically in Fig. \ref{fig:singularities}.

We note that the singularities $q_-=\pm \pi- 2\I L +O(\I L^3)  $ are located in $q\in \mathbb{C}^-$ but they are invisible for any function $f(q)$ which is $2\pi$  periodic because the jumps at the corresponding branch cut $q\in(\pm \pi- 2\I L +O(\I L^3),\pm\pi-\I \infty)$ are identically zero, see also Refs.
\cite{DyachenkoLushnikovKorotkevichPartIStudApplMath2016,LushnikovStokesParIIJFM2016}
for somewhat similar discussion. Because of that we show points $q_-$ by filled circles for the transformation in right panel of Fig. \ref{fig:w-and-q} but do not show $q_-$ in right panel of Fig.   \ref{fig:singularities}.
 If instead of a Stokes wave one would consider a function $\tilde z(w)$ with the branch cut of finite extent then the end point of the mapping of that branch cut into $q$ plane would be not $q_+$ but other point located higher above the real axis. However we do not consider such functions in this paper.

 The singularity $q_{+}$ dominates the asymptotic of Stokes wave FT in $q$ variable provided $|Im(q_{+})|<q_c$. Thus the best convergence of FT (faster decays of Fourier harmonics at large $k)$ occurs for  $|Im(q_{+})|=q_c$ which together with Eqs. \e{qcdefser} and \e{quzeroser} give the optimal choice of the parameter $L=L_{optimal}$ given by Eq. \e{Loptimal} and valid for $v_c\ll 1.$ For $L=L_{optimal}$ both singularities of Stokes wave in $q$ space are located at a distance $\approx(2v_c)^{1/2}$ from the real axis ensuring the FT asymptotic \e{vc1p2}.

\section{Equation of Stokes wave }
\label{sec:StokesWave}

The closed equation for Stokes wave has the following form

\begin{equation}\label{stokes_wave2}
\begin{split}
&   \left(  {c^2}\hat k_u - 1 \right) y -  \left( \frac{\hat k_u y^2}{2} + y\hat k_u y \right) = 0,
%& c_0^2=\frac{g\lambda}{2\upi},
\end{split}
\end{equation}
which is defined on the real line $w=u$ for the function $y(u)$
which is the imaginary part of Eq. \e{zwdef3}. Eq. \e{stokes_wave2}
was derived in Ref. \cite{BabenkoSovietMathDoklady1987} and later
was independently obtained from results of Ref. \cite{DKSZ1996} in
Ref.  \cite{DyachenkoLushnikovKorotkevichJETPLett2014} from the exact Euler equations of free surface
hydrodynamics. See also Ref. \cite{ZakharovDyachenkoPhysD1996} for
somewhat similar equation. Here  $\hat k$ is the positive-definite
linear operator defined by      $\hat k_u := -\frac{\partial}{\p u}
\hat H = \sqrt{-\frac{\partial^2}{\p u^2}}$ and $\hat H_u$ is the
Hilbert transform,
\begin{equation} \label{Hilbertdef}
\hat H _uf(u)=\frac{1}{\pi} \text{p.v.}
\int\limits^{+\infty}_{-\infty}\frac{f(u')}{u'-u}\D u'
\end{equation}
with $\text{p.v.}$ designating a Cauchy principal value of integral and subscript in $u$ means that both the Hilbert transform and $\hat k_u$ are defined for the variable $u$.
The Hilbert operator $\hat H_u$ is a multiplication operator on the Fourier coefficients:
\begin{equation} \label{Hfk}
 (\hat H_u f)_k=\I\,
\text{sign}{\,(k)}\,f_k,
\end{equation}
where $f_k$ are the Fourier coefficients (harmonics):
\begin{align} \label{ffourier}
f_k=\frac{1}{2\pi}\int\limits_{-\pi}^{\pi} f(u)\exp\left (-\I
ku\right )\D u,
\end{align}
 of  the periodic function $f(u)=f(u+2\pi)$ represented through the Fourier series
\begin{equation} \label{fkseries}
f(u)=\sum\limits_{k=-\infty}^{\infty} f_k\exp\left (\I
ku\right ).
\end{equation}
Here $\text{sign}(k)=-1,0,1$ for $k<0, \ k=0$ and $k>0$, respectively.

After solving Eq. \e{stokes_wave2} numerically as described in Refs.   \cite{DyachenkoLushnikovKorotkevichJETPLett2014,DyachenkoLushnikovKorotkevichPartIStudApplMath2016}, we recover the real part $x(u)$ of Eq. \e{zwdef2} from $y(u)$ as
\begin{align} \label{xytransform}
x=u-\hat H_u y,
\end{align}
which follows from the analyticity of $\tilde{z}(w)$  \e{zwdef3}
in $\mathbb{C}^-$ (see a derivation of Eq. \e{xytransform} e.g. in Ref. \cite{DyachenkoLushnikovKorotkevichPartIStudApplMath2016}). Then Stokes wave solution is represented in the parametric form $(x(u),y(u)).$

Eq. \e{stokes_wave2}  was derived in Ref.
\cite{DyachenkoLushnikovKorotkevichJETPLett2014}
under the assumption that  %
\begin{equation} \label{yxucondition}
\int\limits^{\pi}_{-\pi}\eta(x)\D x=\int\limits^{\pi}_{-\pi} y(u)x_u(u)\D u=\int\limits^{\pi}_{-\pi} y(u)[1+\tilde x_u(u)]\D u=0,
\end{equation}
meaning that the mean elevation of  the free surface
is set to zero. Equation \e{yxucondition}  reflects a conservation of the total mass of fluid.

In this paper instead of solving Eq. \e{stokes_wave2} in $u$-variable, we transform it into $q$-variable using  \e{unewdef}. Then we solve the resulting equation numerically in a more efficient way using the procedure described in Section \ref{sec:stokeswavecomputation}. We express $u$ through $q$ as given by Eq. \e{unewdef}, and obtain from Eqs. \e{stokes_wave2} and \e{qjacobian} that
\begin{align}\label{stokes_wave2q}
\left( {c^2} q_u \hat k_q - 1 \right) y  -  \left(q_u \frac{\hat k_qy^2}{2} + q_u\hat y\hat k_q y \right) = 0,
%&   \left(  {c^2} \frac{1+L^2+(1-L^2)\cos{q}}{2L}\hat k_q - 1 \right) y \nonumber \\&-  \left( {} \frac{1+L^2+(1-L^2)\cos{q}}{2L} \frac{\hat k_qy^2}{2} + %\frac{1+L^2+(1-L^2)\cos{q}}{2L}\hat  y\hat k_q y \right) = 0,
%& c_0^2=\frac{g\lambda}{2\upi},
\end{align}
where the  operators
\begin{equation}\label{kqdef}
\hat k_q:= -\frac{\partial}{\p q} \hat H_q =-\frac{\partial}{\p q}
(\hat H_q + const) \end{equation}
 and
\begin{equation} \label{Hilbertdefq}
\hat H _qf(q)=\frac{1}{\pi} \text{p.v.} \int\limits^{+\infty}_{-\infty}\frac{f(q')}{q'-q} \D q'
\end{equation}
now act in $q$ space with $q_u$ given by Eq.  \e{qjacobian}.  Here and below we abuse notation and use the same symbol $y$ for both functions of  $u$ and $q$  (in other words, we  assume that $\tilde y(q)= y(u) $ and remove $\tilde ~$ sign). The comparison of Eqs. \e{stokes_wave2} and \e{stokes_wave2q} together with Eqs. \e{Hilbertdef} and \e{Hilbertdefq} reveals that we simply replaced  $\hat H_u$ by  $\hat H_q  + const$, where the explicit expression for a constant is not important for solving Eq.  \e{stokes_wave2q} because it includes derivatives over $q$ thus removing this constant.  The justification of the validity of this nontrivial replacement is provided in Section \ref{sec:ProjectorsHilberttransformationq}. We also note by comparison of the definitions of $\hat k_u$ and $\hat k_q$ above in this Section that FT of  $\hat k_q$  has the same  meaning of the multiplication on $|k|$ but this time in Fourier space of $q$.

Because the Jacobian $q_u$ is nonzero for any real values $q\in[-\pi,\pi],$ one immediately obtains from equation \e{stokes_wave2q}  a more compact expression
\begin{equation}\label{stokes_wave2q2}
%\left(  {c^2} \hat k _q- \frac{2L}{1+L^2+(1-L^2)\cos{q}} \right) y -  \left( {}  \frac{\hat k _qy^2}{2} +{} y\hat k_q y \right) = 0.
\left(  {c^2} \hat k _q -  \frac{2L}{1+L^2+(1-L^2)\cos{q}} \right) y -  \left( {}  \frac{\hat k _qy^2}{2} +{} y\hat k_q y \right) = 0,
\end{equation}
which we use for simulations.
The mean level zero condition \e{yxucondition} is transformed to
\begin{equation}\label{mean_levelq}
\int\limits_{-\pi}^{\pi}y(q)[u_q+\tilde{x}_q(q)]dq=0
\end{equation}
in the $q$ variable.

\section{Projectors and Hilbert transformation in $q$ variable}
\label{sec:ProjectorsHilberttransformationq}
In this Section we justify the use of the operator $\hat k_q$ in Eqs.  \e{stokes_wave2q} and  \e{stokes_wave2q2}.
It is convenient to introduce the  operators
\begin{equation} \label{Projectordef}
\hat P_u^-=\frac{1}{2}(1+\I \hat H_u)  \quad\text{and}\quad  \hat P_u^+=\frac{1}{2}(1-\I \hat H_u)
\end{equation}
which are the projector operators of a general periodic function into a functions analytic in $w\in\mathbb{C}^-$ and $w\in\mathbb{C}^+$ correspondingly.  To understand the action of these projector operators, we introduce the splitting of  a  general $2\pi$ periodic function $f(u)$ with the Fourier series \e{fkseries} as
\begin{equation} \label{fpm0}
f(u)=f^+(u)+f^-(u)+f_{0,u},
\end{equation}
where
\begin{equation} \label{fplus}
f^+(u)=\sum\limits_{k=1}^{\infty} f_k\exp\left (\I
ku\right )
\end{equation}
is the analytical (holomorphic) function  in $\mathbb{C}^+$ and
\begin{equation} \label{fminus}
f^-(u)=\sum\limits_{k=-\infty}^{-1} f_k\exp\left (\I
ku\right )
\end{equation}
is the analytical function  in $\mathbb{C}^-$ as well as $f_{0,u}$ is the zero Fourier harmonic defined through Eq. \e{ffourier} as
$f_{0,u}=f_k|_{k=0}=\frac{1}{2\pi}\int\limits_{-\pi}^{\pi} f(u ) du$. Together with the property
\begin{equation} \label{Hfpm}
\hat H_u f=\I[f^+(u)-f^-(u)]
\end{equation}
which follows from Eq. \e{Hfk} we obtain that
\begin{equation} \label{Pfm}
\hat P_u^-f=\frac{f_{0,u}}{2}+f^-(u) \quad\text{and}\quad \hat P_u^+f=\frac{f_{0,u}}{2}+f^+(u),
\end{equation}
i.e. the functions which are holomorphic in $\mathbb{C}^-$ and $\mathbb{C}^+$, respectively.

Here we use the notation  $\hat P_u^{-}$, $\hat P_u^{+}$ and  $\hat H_u$ for the projectors and Hilbert transform in $u$ space.
%In a similar way,   $\hat P_q^{-}$, $\hat P_q^{+}$  and  $\hat H_q$  are for $q$ space.
Similarly, we introduce projector operators $\hat P_q^{-}$, $\hat P_q^{+}$  and Hilbert transform $\hat H_q$ in the $q$-variable.
%To find how the projectors $\hat P_u^{-}$, $\hat P_u^{+}$ and Hilbert transform $\hat H_u$ look like in $q$ space one can use two approaches.
We can use two approaches to determine the form of projectors $\hat P_u^{-}$, $\hat P_u^{+}$ in the $q$ space.
The first approach is to analyze how Fourier series transforms as we make a change of variables from $u$ to $q$. The second approach is to use definition of these operators through complex contour integrals and see how these integrals transform  as we make a change of variables from $u$ to $q$. In this paper we focus on the second approach as well as we provide the expressions only for   $\hat P_q^{-}$ and  $\hat H_q$. The expression for    $\hat P_q^{+}$ can be derived in a similar way but it is not need for the computation of Stokes wave.

Using  the Sokhotskii-Plemelj theorem  (see e.g. \cite{Gakhov1966,PolyaninManzhirov2008})
\begin{align} \label{sokhotskii}
 \int\limits^{\infty}_{-\infty}\frac{f(u')du'}{u'-u+\I 0}=\text{p.v.}\int\limits^{\infty}_{-\infty}\frac{f(u')du'}{u'-u}-\I \pi  f(u),
\end{align}
where $\I 0$ means $\I \epsilon, \ \epsilon\to 0^+$, we rewrite Eq. \e{Projectordef} as follows
\begin{align} \label{Puminus}
\hat P_u^-f=\frac{1}{2}(\I \hat H_u+1)f=-\frac{1}{2\pi \I}\text{p.v.}\int\limits^{\infty}_{-\infty}\frac{f(u')du'}{u'-u}+\frac{1}{2}f(u)=-\frac{1}{2\pi \I}\int\limits^{\infty}_{-\infty}\frac{f(u')du'}{u'-u+\I 0}.
\end{align}
We now use $2\pi$ periodicity of $f(u)$ to reduce Eq.  \e{Puminus} into the integral over one period
\begin{align} \label{Puminusperiod}
\hat P_u^-f=-\frac{1}{2\pi \I}\sum \limits _{n=-\infty}^\infty \int\limits^{\pi}_{-\pi}\frac{f(u')du'}{u'-u+\I  0+2\pi n}=-\frac{1}{4\pi \I} \int\limits^{\pi}_{-\pi}\frac{f(u')du'}{\tan{\frac{u'-u+\I  0}{2}}}.
\end{align}
Using equations \e{unewdef} and \e{qjacobian} we transform Eq. \e{Puminusperiod} to $q$-variable as follows
\begin{align} \label{Puminusperiodq}
&\hat P_u^-f=-\frac{1}{4\pi \I} \int\limits^{\pi}_{-\pi}\frac{f(q')dq'}{\tan{\frac{u'-u+\I  0}{2}}}u_{q'}dq'
\nonumber \\&=-\frac{1}{4\pi \I} \int\limits^{\pi}_{-\pi}\frac{f(q')\left [1+L^2\tan{\frac{q'+\I  0}{2}}\tan{\frac{q  }{2}}\right ]dq'}{\tan{\frac{q'+\I  0}{2}}-\tan{\frac{q  }{2}}}\frac{1}{\cos^2{\frac{q'}{2}}\,
\left(1+L^2\tan^2{\frac{q'}{2}}\right )}.
\end{align}

Similar to Eq. \e{fpm0} we write $f(q)$ in $q$ space as follows
\begin{equation} \label{fpm0q}
f(u)\equiv f(q)=f^{+,q}(q)+f^{-,q}(q)+f_{0,q},
\end{equation}
where
\begin{equation} \label{fplusq}
f^{+,q}(q)=\sum\limits_{k=1}^{\infty} f_k\exp\left (\I
kq\right )
\end{equation}
is analytic function in $\mathbb{C}^+$ and
\begin{equation} \label{fminus}
f^{-,q}(q)=\sum\limits_{k=-\infty}^{-1} f_k\exp\left (\I
kq\right )
\end{equation}
is the analytical function  in $\mathbb{C}^-$ and $f_{0,q}$ is the
zero Fourier harmonic
$
f_{0,q}=\frac{1}{2\pi}\int\limits_{-\pi}^{\pi} f(q) dq.
$

% w and q plane figure
\begin{figure}
\includegraphics[width = 0.99\textwidth]{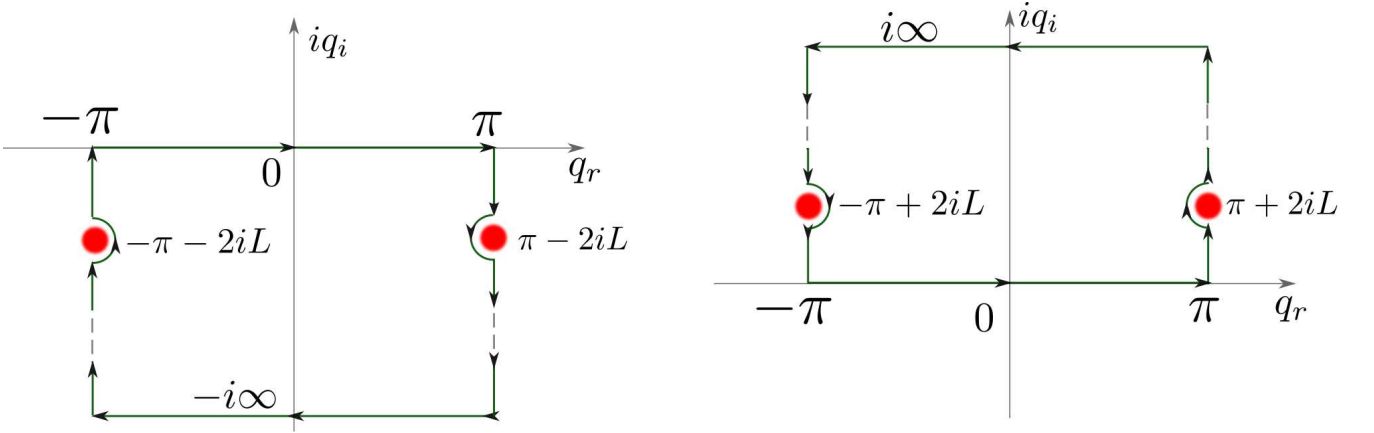}
\caption{A schematic of the integration contours in $q\in
\mathbb{ C}^-$ (left panel) and  $q\in \mathbb{C}^+$ (right panel)
used for evaluating different parts of the integral in Eq. \e{Puminusperiodq}. These
contours bypass from inside  (by pairs of  infinitesimal half-circles)
the singularities \e{quzero} of the conformal map  \e{unewdef}  at
points   $q=\pm2 \arctan(\I/ L)\pm \pi=\pm2\I L\pm\pi +O(\I L^3)   $
(shown by filled circles).  Vertical parts of contours are canceled
out during integration by the periodicity  of  the integrand.
 }
\label{fig:contour}
\end{figure}

We evaluate integrals in  Eq. \e{Puminusperiodq} using \e{fpm0q} by closing complex contours in $q\in \mathbb{C}^+$ for
$f^+(q)$ and   $q\in \mathbb{C}^-$ for  $f^-(q),$ respectively as shown in Fig. \ref{fig:contour}. For $f_{0,q}$ it can be done in both ways giving the same result. The zeros of the denominator are located at $q'=q-\I 0$ and $q'=\pm 2\arctan{\frac{\I}{L}}$. We calculate the residues to obtain:
\begin{equation}\label{Puplusperiodq2b}
\begin{split}
&\hat P_u^{-}f^{+,q}(q)=-\frac{2\pi\I}{4\pi \I}
\frac{f^{+,q}(2\arctan{\frac{\I}{L}})\left
[1+L^2\frac{\I}{L}\tan{\frac{q  }{2}}\right
]}{\frac{\I}{L}-\tan{\frac{q  }{2}}}\frac{1}{
 \I L}=\frac{1}{2}f^{+,q}\left(2\arctan{\frac{\I}{L}}\right),  \\
&\hat P_u^{-}f^{-,q}(q)=%(-1)^2
\frac{2\pi\I}{4\pi \I}
\frac{f^{-,q}(q)\left [1+L^2\tan{\frac{q}{2}}\tan{\frac{q
}{2}}\right
]}{\frac{1}{2\cos^2{\frac{q}{2}}}}\frac{1}{\cos^2{\frac{q}{2}}\,
\left(1+L^2\tan^2{\frac{q}{2}}\right )}%\nonumber
\\&+%(-1)^2
\frac{2\pi\I}{4\pi \I}
\frac{f^{-,q}(-2\arctan{\frac{\I}{L}})\left
[1+L^2\frac{-\I}{L}\tan{\frac{q  }{2}}\right
]}{\frac{-\I}{L}-\tan{\frac{q  }{2}}}\frac{1}{
 -\I
 L}=f^{-,q}(q)-\frac{1}{2}f^{-,q}\left(-2\arctan{\frac{\I}{L}}\right),
 \\
&\hat P_u^{-}f_{0,q}=\frac{1}{2}f_{0,q}.
\end{split}
\end{equation}
Using equations  \e{fpm0q}  and \e{Puplusperiodq2b} yields:
\begin{align} \label{Pminusq3}
\hat P_u^-f(q)=\frac{f_{0,q}}{2}+f^{-,q}(q)-\frac{1}{2}f^{-,q}\left(-2\arctan{\frac{\I}{L}}\right)+\frac{1}{2}f^{+,q}\left(2\arctan{\frac{\I}{L}}\right).
\end{align}
Defining the projector $\hat P^-_q$ in $q$ space similar to Eq. \e{Pfm} as
\begin{equation} \label{Pfmq}
\hat P_q^-f=\frac{f_{0,q}}{2}+f^-(q),
\end{equation}
we obtain from Eq. \e{Pminusq3} that
\begin{align} \label{Pminusq3a}
\hat P_u^-f(q)=\hat P_q^-f+c_{shift},
\end{align}
where
\begin{equation} \label{cdef}
c_{shift}:=-\frac{1}{2}f^{-,q}\left(-2\arctan{\frac{\I}{L}}
\right)+\frac{1}{2}f^{+,q}\left(2\arctan{\frac{\I}{L}}\right)
\end{equation}
is the constant.
We define $\hat H_q$ through a relation similar to Eq. \e{Projectordef} in $q$ space
\begin{equation} \label{Projectordefq}
\hat P_q^-=\frac{1}{2}(1+\I \hat H_q),
\end{equation}
and we obtain from Eq. \e{Pminusq3a} and \e{Projectordefq} that
\begin{align} \label{Hfpmq}
\hat H_u^-f(q)=\hat H_q^-f -2\I c_{shift}.
\end{align}
Thus the operators $\hat P^-$ and $\hat H$ in $u$ and
$q$ spaces are the same except for the shift by a constant
$c_{shift}$ and $-2\I c_{shift}$ respectively.  These
constants result from the singularities  \e{quzero} of the conformal
map \e{unewdef}.  The explicit expression for $c_{shift}$ is
calculated from the values of $f^{-,q}(q)$ as follows. We  notice that for Stokes wave $y(u)$ is an even real function, $y(u)=y(-u)\in\mathbb{R},$ which implies that    $y(q)=y(-q)\in\mathbb{R}$ in $q$ variable. By taking $f(q)=y(q)$ we obtain that $f^{+,q}(q)= f^{-,q}(-q)$. The analytical continuation of  $f^{-,q}(q)$ from the real line $q=Re(q)$ into the complex value $q=-2\arctan{\frac{\I}{L}}$ is trivially done  by plugging the complex value of $q$ into the series \e{fminus} reducing Eq. \e{cdef} to
\begin{equation} \label{cval}
c_{shift}=0
\end{equation}
for the even real function $y(q).$ For more general non-even solution of Eq. \e{stokes_wave2q2} (corresponds to higher order progressive waves, which
have more than one different peaks per $2\pi$ spatial period \cite{ChenSaffmanStudApplMath1980}) we  generally obtain nonzero value of $c_{shift}$ by a similar procedure as follows.  We recover $f^{+,q}(q)$  from  $f^{-,q}(q)$ using the identity
\begin{equation} \label{fpm1}
f^{+,q}(q)= \bar f^{-,q}(-q),
\end{equation}
which follows from the condition that $f(u)$ is the real-valued function. Here $\bar f(q)$ means the complex conjugation of  the function $f(q)$ for real values of $q$, i.e.  $\bar f(q)\equiv \overline {f(\bar{q})}$ for complex values of $q.$ For Eq. \e{fminus}  it implies that  $\bar f^{-,q}(q)=\sum\limits_{k=-\infty}^{-1} \bar f_k\exp\left (-\I
kq\right ).$ Then $c_{shift}$
results from the analytical continuation of  $f^{-,q}(q)$ and $\bar f^{-,q}(-q)$ from the real line into  $q=-2\arctan{\frac{\I}{L}}$ together with Eqs. \e{cdef} and  \e{fpm1}.

 Note
that we do not need the explicit  value for $c_{shift}$ to solve  Eqs.
\e{stokes_wave2q} and  \e{stokes_wave2q2} because they both include derivatives over
$q$ which removes $c_{shift}$. However, to obtain $x(u)$  one generally needs the value of $c_{shift}$ (which  produces only a trivial shift in the horizontal direction). Using Eqs.  \e{unewdef}, \e{Hfpmq}, \e{cdef}, we transform Eq. \e{xytransform}  into the variable $q$ as follows
\begin{align} \label{xytransformq}
x=u(q)-\hat H_u y=u(q)-\hat H_q y-2\I c_{shift}.
\end{align}
We conclude that this section has  justified the derivation
of Eqs.  \e{stokes_wave2q} and  \e{stokes_wave2q2} from Eq.
\e{stokes_wave2}.

\section{Numerical algorithm for computing Stokes wave}
\label{sec:stokeswavecomputation}

We solve Eq. \e{stokes_wave2q2} numerically using the generalized Petviashvili method (GPM) \cite{LY2007,PelinovskyStepanyantsSIAMNumerAnal2004} and the Newton Conjugate Gradient method proposed in Refs.~\cite{JiankeYang2009,YangBook2010}. These numerical methods are similar to the numerical solution of Eq.
  \eqref{stokes_wave2} in Refs.   \cite{DyachenkoLushnikovKorotkevichJETPLett2014,DyachenkoLushnikovKorotkevichPartIStudApplMath2016}.
For both methods  $y(q) $ is expanded in
cosine Fourier series and the operator $\hat k_q$ \e{kqdef} is evaluated
numerically using Fast Fourier Transform (FFT) on the uniform grid with
$M_q$ points discretization of the interval $-\pi\le q< \pi$.

As alternative to solving Eq. \e{stokes_wave2q2}, we also numerically solved the equivalent equation
\begin{equation}\label{stokes_wave3q}
c^2\tilde{z}_q =-\I \hat P_q^-[(\tilde z-\bar{{\tilde z}})(u_q +\tilde{z}_q)],
\end{equation}
which is the analog of equation
\begin{equation}\label{stokes_wave3}
c^2\tilde{z}_u = -\I  \hat P_u^-[(\tilde z-\bar{{\tilde z}})(1+\tilde{z}_u)],
\end{equation}
derived in Ref.  \cite{LushnikovStokesParIIJFM2016}. Eq. \e{stokes_wave3} is equivalent to Eq. \e{stokes_wave2}
and is obtained by applying the projector operator  $\hat P_u^-$   \e{Projectordef} to equation   \e{stokes_wave2}
together with the condition
 \e{yxucondition}. In a similar way. Eq. \e{stokes_wave3q} is obtained by applying the projector operator $\hat P_q^-$ \e{Projectordefq} to equation   \e{stokes_wave2q2}
together with the condition
 \e{mean_levelq}.

Solving Eq. \e{stokes_wave3q} numerically instead of  Eq. \e{stokes_wave2q2} typically provides 1-2 extra digits of accuracy in Stokes wave height $H$ as well as in the accuracy of the solution spectrum and $v_c$. The extra cost is however that we have to solve Eq. \e{stokes_wave3q} for the complex-valued function $\tilde z(q)$ instead of the real valued function  $ y(q)  $ in Eq. \e{stokes_wave2q2} which doubles memory requirements and the number of numerical operations.

After we obtain a numerical solution for $z(q),$ we use it to determine the value of $v_c$ via one of three numerical methods:
%\begin{enumerate}
  %\item

(i) The first method uses a least squares fit of Fourier spectrum of a solution $\tilde z$ to the asymptotic series described in Eq.   (41) of Ref. \cite{DyachenkoLushnikovKorotkevichPartIStudApplMath2016}. Working in $u$ variable this method allows one to obtain $v_c$ with the absolute accuracy about $10^{-10}-10^{-11}$ in double precision (DP) using 7-12 terms in the series of Eq.   (41) of Ref. \cite{DyachenkoLushnikovKorotkevichPartIStudApplMath2016}. While working in $q$ variable, the second singularity \e{quzero} located at  $q=q_+= \pm \pi + 2\I L+O(\I L^3) $ introduces a contribution to the Fourier spectrum of the same order as the main singularity \e{qcdef} (located at $q=\I q_c\simeq \I  v_c/L$) if the parameter $L$ is chosen close to $L_{optimal}\simeq(v_c/2)^{1/2}$ \e{Loptimal}, so we typically can get only 1-2 digits of precision in $v_c$. In order to obtain $v_c$ with higher accuracy one needs to remap the solution via Fourier interpolation to a uniform  grid for the new variable $\tilde{q}$ with the larger value of the parameter $\tilde{L}$ (we found that a factor 8 or 16 is typically enough to obtain $v_c$ with maximum possible accuracy in DP). This  pushes the second singularity much further away from the real line compared to the first one so that the main contribution to the tail of Fourier spectrum in $\tilde{q}$ space comes from the first singularity (at a distance $\simeq v_c/\tilde{L}$ from the real line) which will allow us to find $v_c/\tilde{L}$ (and consequently $v_c$) using the same fitting procedure as in $u$ space. Using this approach for solutions in $q$ space we were able to recover $v_c$ with absolute accuracy about $10^{-9}-10^{-10}$ in DP. We typically used it for solution with $M_q<10^5$ Fourier harmonics since Fourier interpolation procedure uses $ O(M_q^2)$ operations and becomes slow for larger $M_q$.

(ii) %  \item
The second method is described in Section 6.1 a of Ref. \cite{LushnikovStokesParIIJFM2016} and based on the compatibility of the series expansions
  at points $\zeta=\pm \I\chi_c$ in the axillary space $\zeta=\tan[\frac{w}{2}]$  (which implies that $\chi_c=\tanh[\frac{v_c}{2}])$  with the equation \e{stokes_wave2}
  of Stokes wave. Current realization of this algorithm also requires  Pad\'e  approximation of the solution in the axillary space $\zeta$ (described in Section
  4 of Ref. \cite{DyachenkoLushnikovKorotkevichPartIStudApplMath2016}) for calculation of coefficients of series expansion at the
  point $\zeta=-\I\chi_c=-\tan[\frac{\I v_c}{2}]$. This method is so far the most accurate but requires $O(M_q\times N_d)$ operations ($N_d$ is the number of poles) for finding Pad\'e  approximation of a solution thus slow for large $M_q$. We typically used that method for
 $M_q>10^5,$ where the the small value of $v_c$ required us to use   quadruple  (quad) precision with 32 digits
accuracy both to obtain $\tilde z$ and recover $v_c.$ The absolute accuracy for $v_c$ in this method was  $\sim 10^{-26}.$

(iii)  %\item
The third method is described in Section 4.3 of Ref. \cite{DyachenkoLushnikovKorotkevichPartIStudApplMath2016} and uses nonlinear fit of the crest of a solution to a series (4.13) in Ref. \cite{DyachenkoLushnikovKorotkevichPartIStudApplMath2016}. One can work in either $q$ space to find $v_c/L$ or $u$ space to find directly $v_c$. The method was used as the substitute of the method (ii) for the smallest values $v_c\lesssim 10^{-10}$ we achieved, where  Pad\'e  approximation become computationally challenging taking more computer time than the calculation of $\tilde z$ itself. The absolute accuracy of that method for $v_c$ is $\sim 10^{-20}.$
%\end{enumerate}

To summarize, we used the first two methods for finding $v_c$ for solution with $v_c>10^{-8}$ in DP, the second method for solutions with with $v_c\gtrsim10^{-10}$ in quad precision and the third method  for solution with $v_c\lesssim10^{-10}$ (with $M_q\gtrsim10^6$). We also performed a multiprecision simulations with a  variable precision arithmetics with  $\sim 200$  digits for selected values of parameters as described in the next Section.

\begin{figure}
\includegraphics[width = 0.85\textwidth]{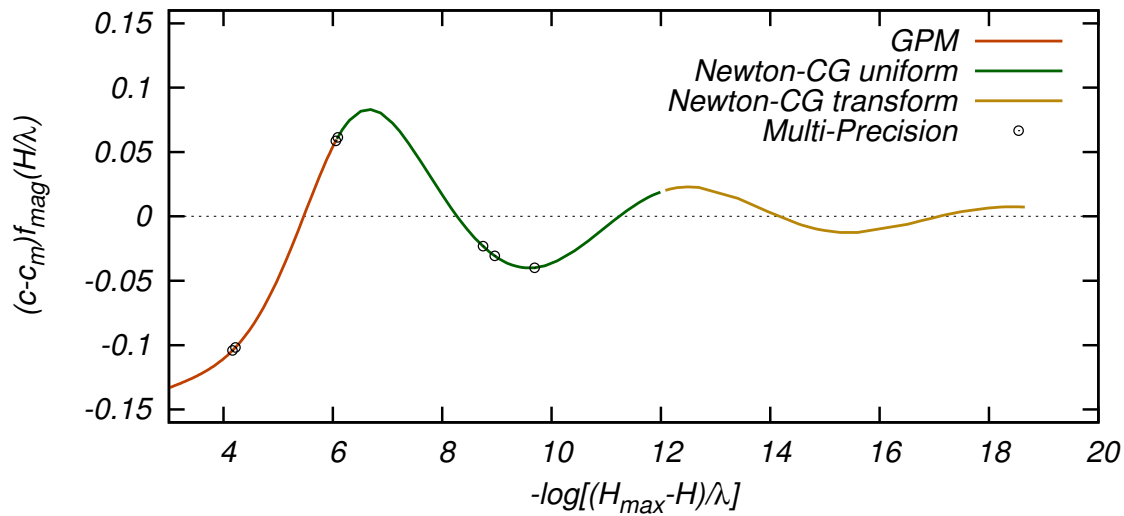}
\caption{Illustration of performance of the numerical methods applied to Eq. ~\eqref{stokes_wave2q2}: (red) waves accesible by means of Generalized Petviashvili method (GPM),
(green) waves accesible via Newton--Conjugate Gradient method (Newton-CG) on a uniform grid $u$, and (gold) %???)
 waves accessible via Newton-CG method on nonuniform
grid with the new conformal  map  \e{qnewdef}. Open circles mark positions of selected Stokes waves separately computed in multiprecision using {\it Wolfram Mathematica} software. The plot is scaled by a magnification function $f_{mag}(H/\lambda) = 1/(30(H_{max}-H)/\lambda)^{1.15} + 1$ to show all simulation data in a single graph while stressing obtained oscillations.}
\label{fig:methods}
\end{figure}

\section{Results of numerical simulations}
\label{sec:results}

Previous results summarized  in Fig. 2 of Ref. \cite{DyachenkoLushnikovKorotkevichPartIStudApplMath2016}  (they are also reproduced in the left part of the curve of Fig.~\ref{fig:methods})  showed a nontrivial dependence of the Stokes wave speed $c$ on the height $H$
with $H$ monotonically approaching the maximum value $H_{max}$ and $c$ approaching a finite value non-monotonically while oscillating with an amplitude that
decreases approximately two orders in magnitude every half of such oscillation. Computing Stokes wave solutions in $u$ space as in
Ref. \cite{DyachenkoLushnikovKorotkevichPartIStudApplMath2016} allowed to resolve about 1.5 of such oscillations (see   Fig.~\ref{fig:methods}) while implementing
the approach described in this paper (solving in $q$ space) allowed to resolve about 3.5 of such oscillations.

\begin{figure}
\includegraphics[width = 0.495\textwidth]{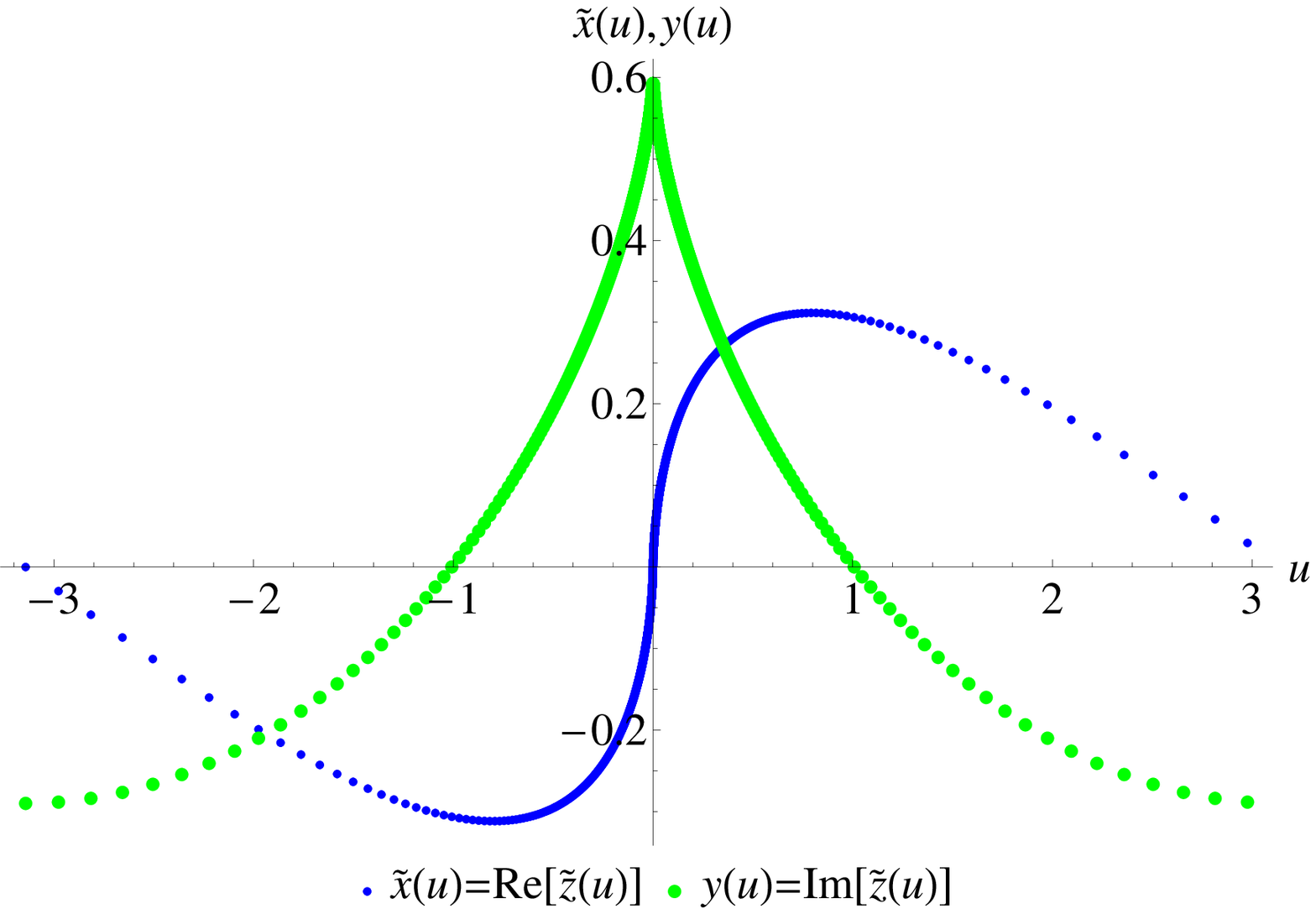}
\includegraphics[width = 0.495\textwidth]{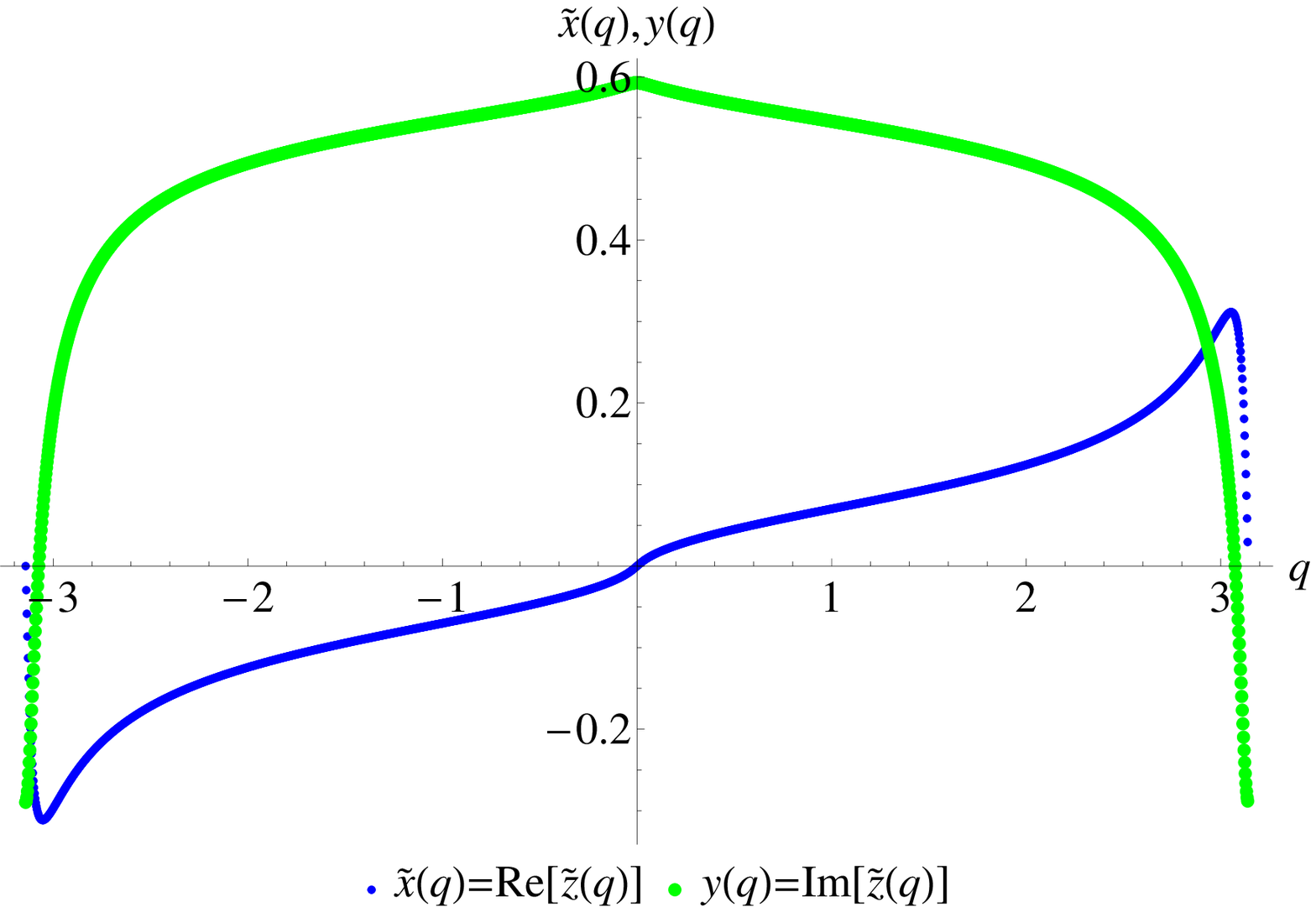}
\caption{Real and imaginary parts of the Stokes wave solution
$\tilde{z}=\tilde{x}+iy$ for  $H/\lambda=0.1404429731116977$ and $c=1.0924$ in $u$ (left) and $q$
(right) variables. It is seen that the gradients of $\tilde z$ are significantly reduced in $q$ variable.  } \label{fig:wave}
\end{figure}

\begin{figure}
\includegraphics[width = 0.495\textwidth]{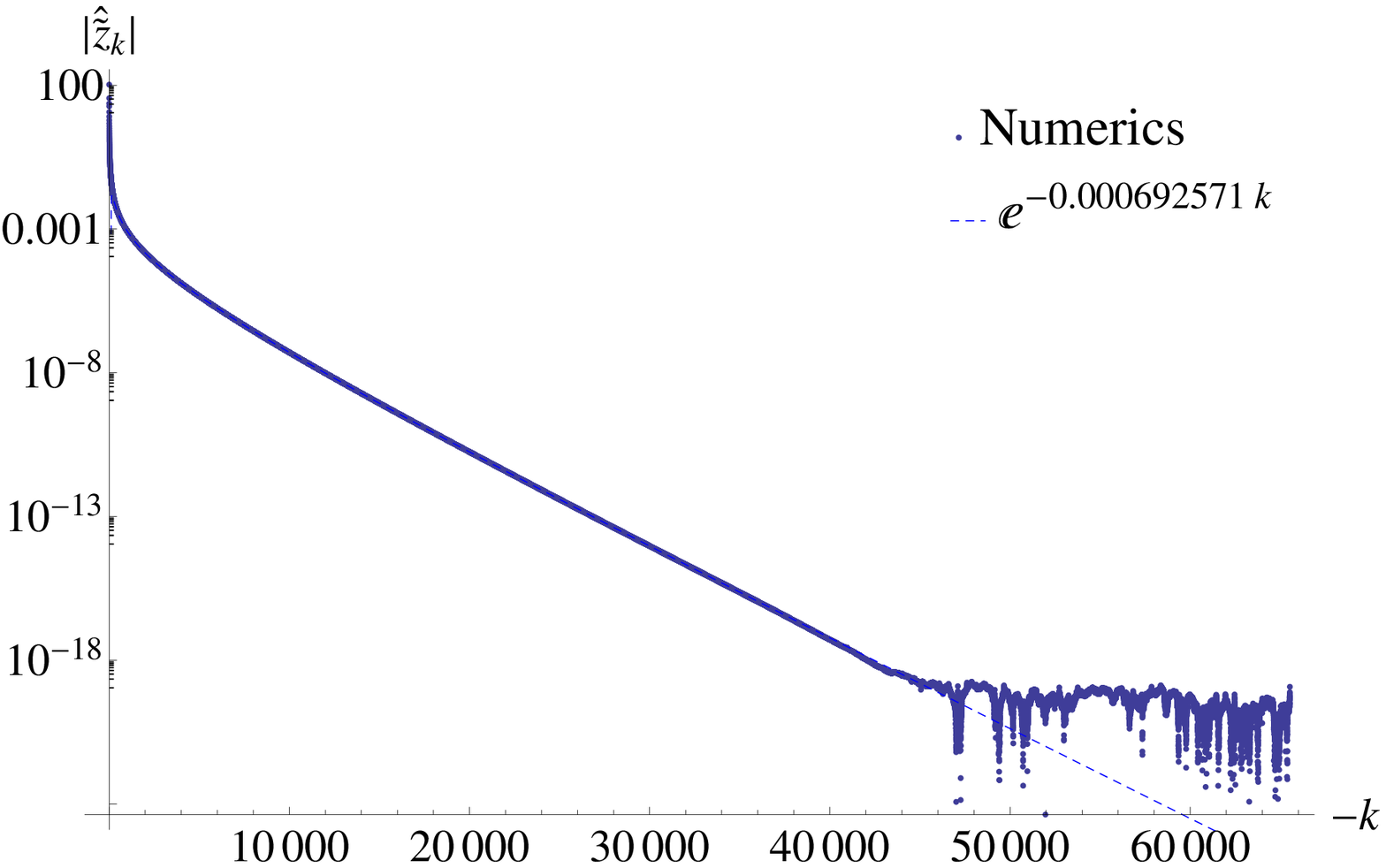}
\includegraphics[width = 0.495\textwidth]{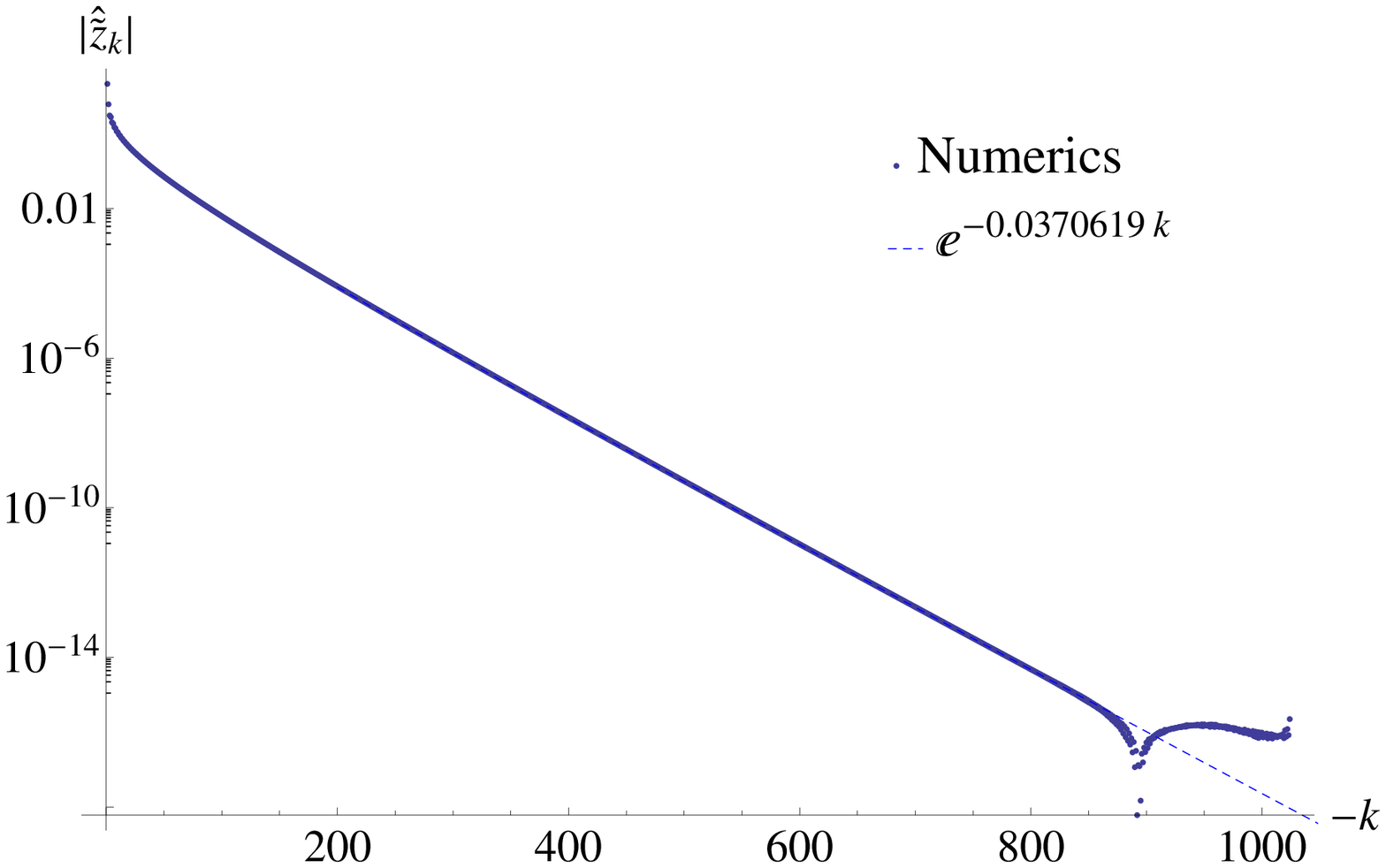}
\caption{Spectra of the Stokes wave for  $\tilde{z}=\tilde{x}+iy$ with  $H/\lambda=0.1404429731116977$, $c=1.0924$ in $w$ (left) and $q$ (right) variables calculated in DP.}
\label{fig:spectra}
\end{figure}

One example of the numerical solution (corresponds to the most extreme wave of Ref. \cite{DyachenkoLushnikovKorotkevichPartIStudApplMath2016})  is given in the Introduction. Another  example of a less  steep wave solution for $c=1.0924$ computed using double precision resulting in $H/\lambda=0.1404429731116977$ and $v_c=0.0006925714\ldots$ is given in Fig. \ref{fig:wave} in variables $u$ (left panel) and $q$ (right panel) with the corresponding spectra of $\tilde{z}(u)$ and $\tilde{z}(q)$ showed in Fig. \ref{fig:spectra} (both spectra have only negative components of $k$ since both  $\tilde{z}(w)$ and $\tilde{z}(q)$ are holomorphic in $\mathbb{C}^-$). Here  $M=64536$ on a uniform grid and $M_{q}=1024$ on a nonuniform grid with $L=0.018608751114420542$.   It demonstrates that for this particular case one needs 64 times less Fourier harmonics in $q$ space compared to the $u$ space in order to resolve the solution up to DP round-off  error. The speed up factor could be roughly estimated as $1/L=(v_c/2)^{1/2}$ that becomes  significant as we go to lower values of $v_c$.

\begin{figure}
\includegraphics[width = 0.495\textwidth]{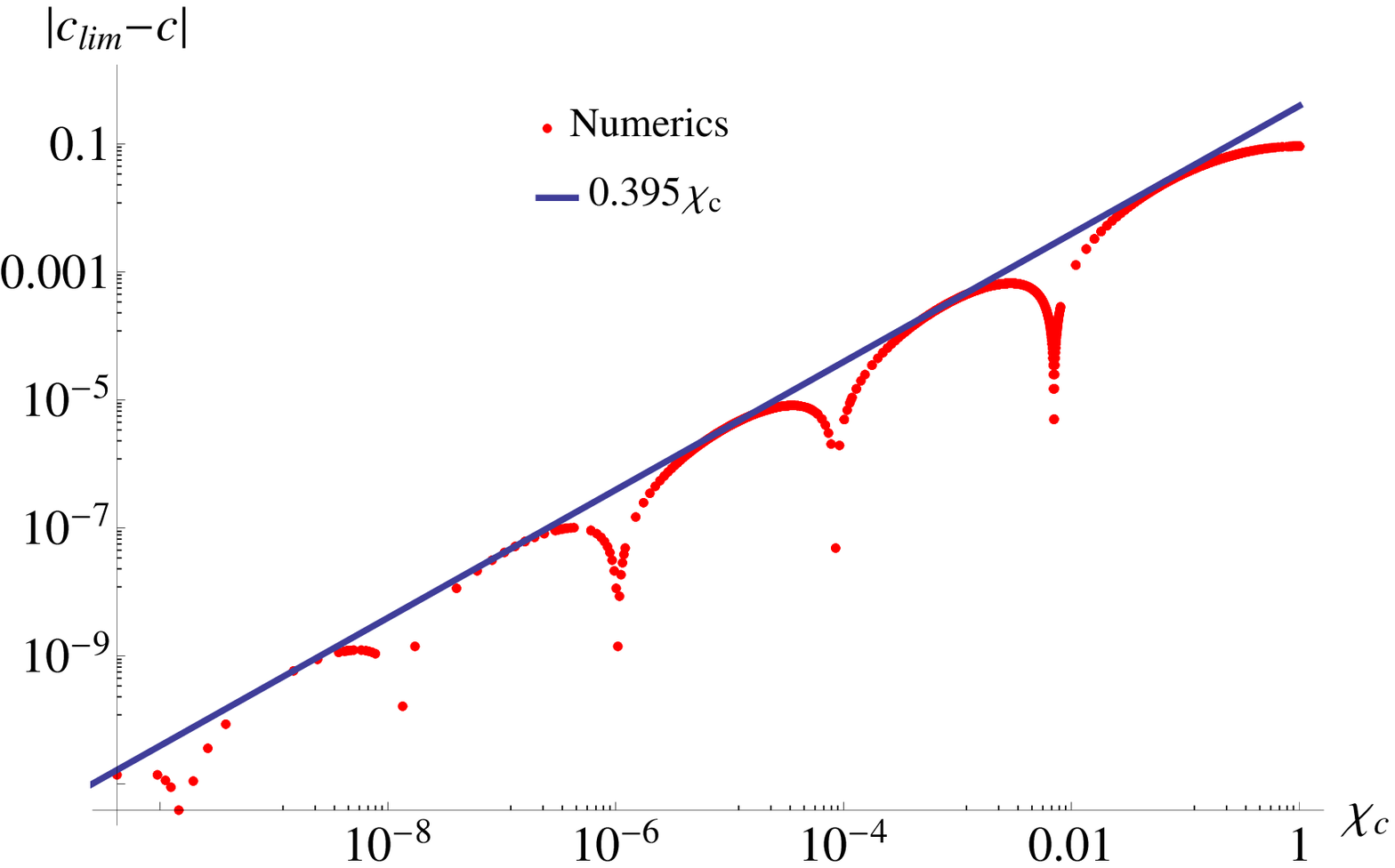}
\includegraphics[width = 0.495\textwidth]{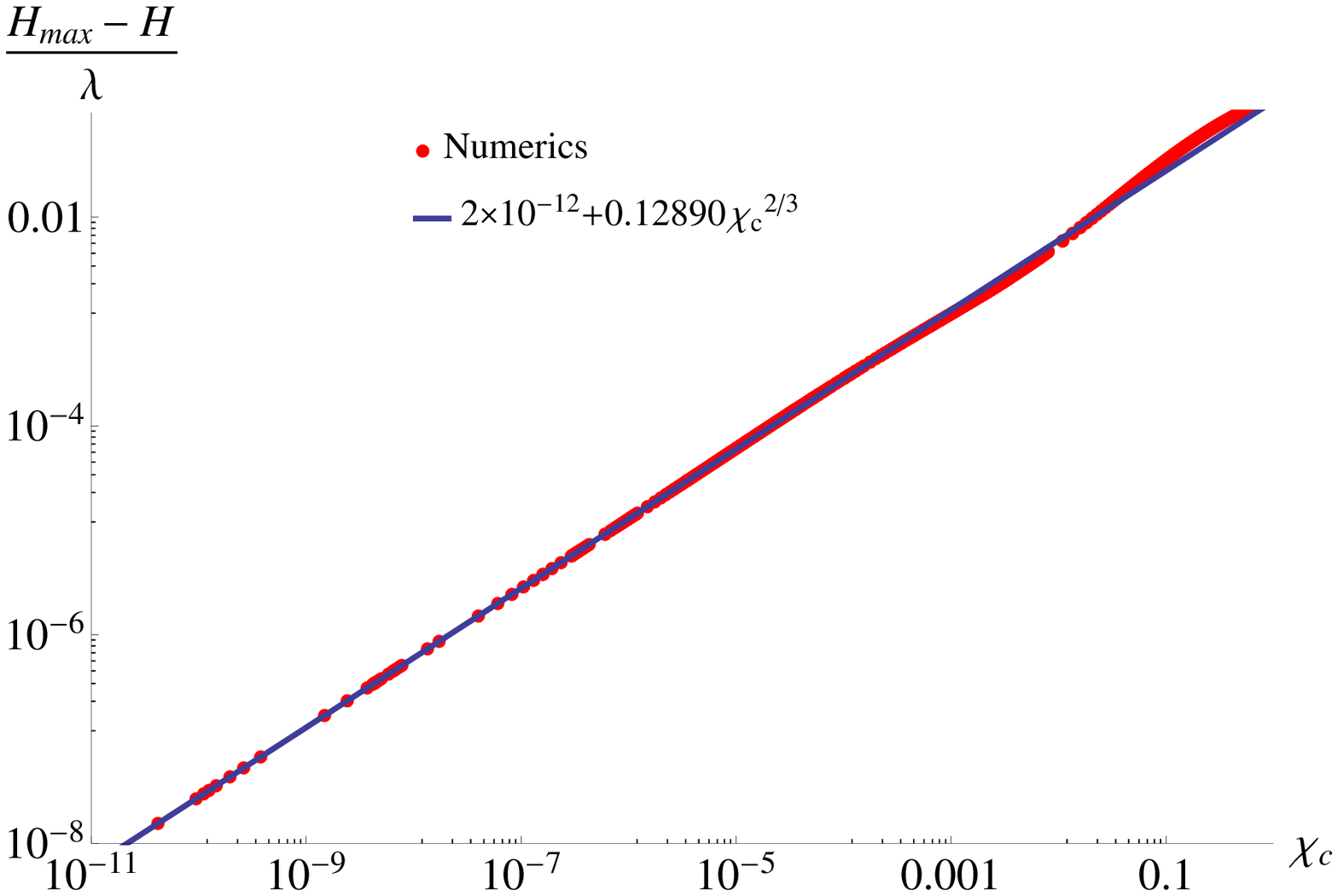}
\caption{ Dependencies of $|c_{lim}-c|$ (left panel) and
$\frac{H_{max}-H}{\lambda}$ (right panel) as  functions of the
parameter $\chi_c$, where $|c_{lim}-c|$ is deviation of the velocity
$c$ of the Stokes wave from the velocity of the limiting Stokes wave
$c_{lim}$ and $\frac{H_{max}-H}{\lambda}$ is the deviation of the
Stokes wave height $H$  the height  $H_{max}$ of the limiting Stokes
wave. The red dots are simulation data  while the solid lines are
their corresponding fits.} \label{fig:cH_vs_Xc}
\end{figure}

High precision and range of our simulation parameters allow to reveal the asymptotic behavior of Stokes wave as it approaches the limiting form as well as make a comparison with the  theory of Stokes wave. We start by analyzing the dependencies of wave speed $c$ and height $H$ on the parameter $\chi_c$ for the obtained family of Stokes waves,
where
\begin{equation} \label{chicdef}
\chi_c=\tanh[\frac{v_c}{2}]
\end{equation}
 is the distance to the singularity of a Stokes wave solution to the real line in the axillary space
$\zeta=\tan[\frac{w}{2}]$. Notice, that for the highly nonlinear Stokes waves  $v_c \rightarrow 0$ and $\chi_c \simeq v_c/2$, while for  the almost linear Stokes waves  $v_c \rightarrow \infty $ and $\chi_c \rightarrow 1$.
Fig. \ref{fig:cH_vs_Xc} shows  $|c_{lim}-c|$ and $(H_{max}-H)/\lambda$ vs. $\chi_c$ for computed Stokes waves in the
log-log scale together with the corresponding fitting curves. Here     $c_{lim}$ and  $H_{max}/\lambda$  are the speed and the scaled height of the limiting Stokes waves, respectively.  We use   the numerical values     $c_{lim} ^{GL}= 1.0922850485861$ and $H^{GL}_{max}/\lambda = 0.1410634839798$
found by I.S. Gandzha and V. P. Lukomsky in Ref. \cite{GandzhaLukomskyProcRoySocLond2007} with the claimed accuracy in 11 digits.
 Comparable accuracy was also achieved in Ref. \cite{MaklakovEuroJnlAppliedMath2002}. It is seen  from Fig. \ref{fig:cH_vs_Xc}  that $|c_{lim}-c|$ experiences oscillations with their envelope being the excellent fit to the linear law  %
\begin{equation} \label{climcscaling}
 Envelope(c_{lim}-c)\propto \chi_c.
\end{equation}
 Fig.  \ref{fig:cH_vs_Xc}.  shows that the dependence of $H_{max}-H$ on $\chi_c$ at the leading order fits well to the the power law
\begin{equation} \label{HmaxHscaling}
H_{max}-H \propto \chi_c^{2/3}
\end{equation}
while experiencing
  small oscillations with the vanishing amplitude as  $\chi_c \rightarrow 0.$    The scaling \e{HmaxHscaling} was proposed in Ref.
\cite{DyachenkoLushnikovKorotkevichJETPLett2014} from simulations and can be extracted at the leading order from the analytical Stokes wave solution of  Section 8 of Ref. \cite{LushnikovStokesParIIJFM2016}. Using the scaling \e{HmaxHscaling}
we fit the simulation data into the model
\begin{equation} \label{HmaxAmodel}
H^{GL}_{max}+\Delta H_{max}-H=\lambda A \chi_c^{2/3},
\end{equation}
 where the constant  $\Delta H_{max}$ accounts for the accuracy in the numerical value of $H^{GL}_{max}$ and $A$ is the another fitting parameter. Using the smallest values of $\chi_c\lesssim 10^{-10}$ achieved in simulations, we obtained from that fit the estimate $H^{fit}_{max}/\lambda =(H^{GL}_{max}+\Delta H_{max})/\lambda= 0.141063483977 \pm 10^{-11},$ i.e. $\Delta H_{max}/\lambda=-2.8\cdot 10^{-12}$  which is consistent with  11 digits accuracy of  $H^{GL}_{max}$. The
highest wave that we computed in QP has $
H_{max}^{lowerbound}/\lambda= 0.1410634805062790\ldots $ (for $c=1.09228504858750000$) which provides the best lower bound  $H_{max}/\lambda $  from our simulations. That lower bound is within $\simeq3.5\times10^{-9}$ from $H^{fit}_{max}/\lambda$ which is more than 3 orders in magnitude of improvement compare with the simulations of Ref. \cite{DyachenkoLushnikovKorotkevichPartIStudApplMath2016}.

\begin{figure}
\includegraphics[width = 0.495\textwidth]{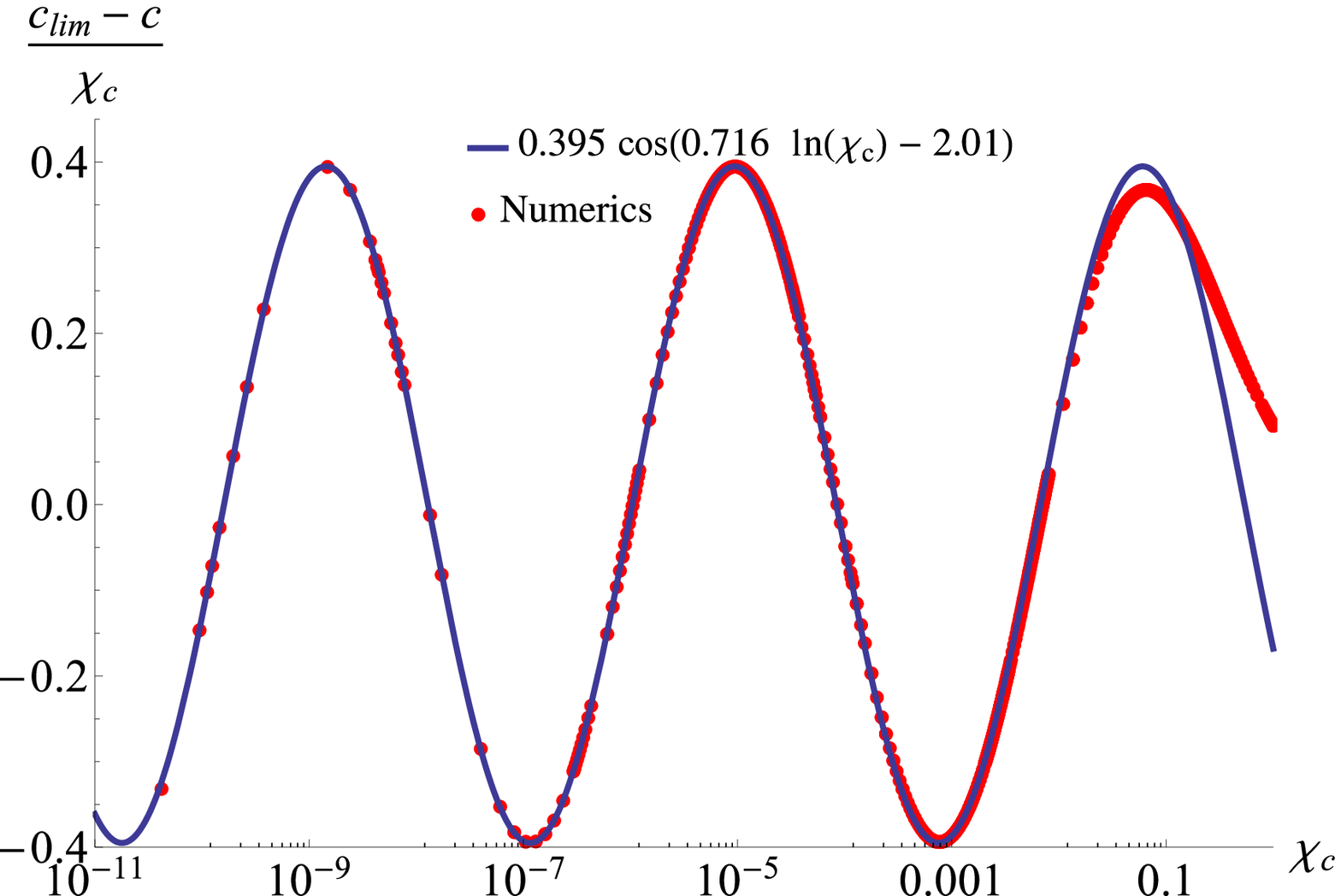}
\includegraphics[width = 0.495\textwidth]{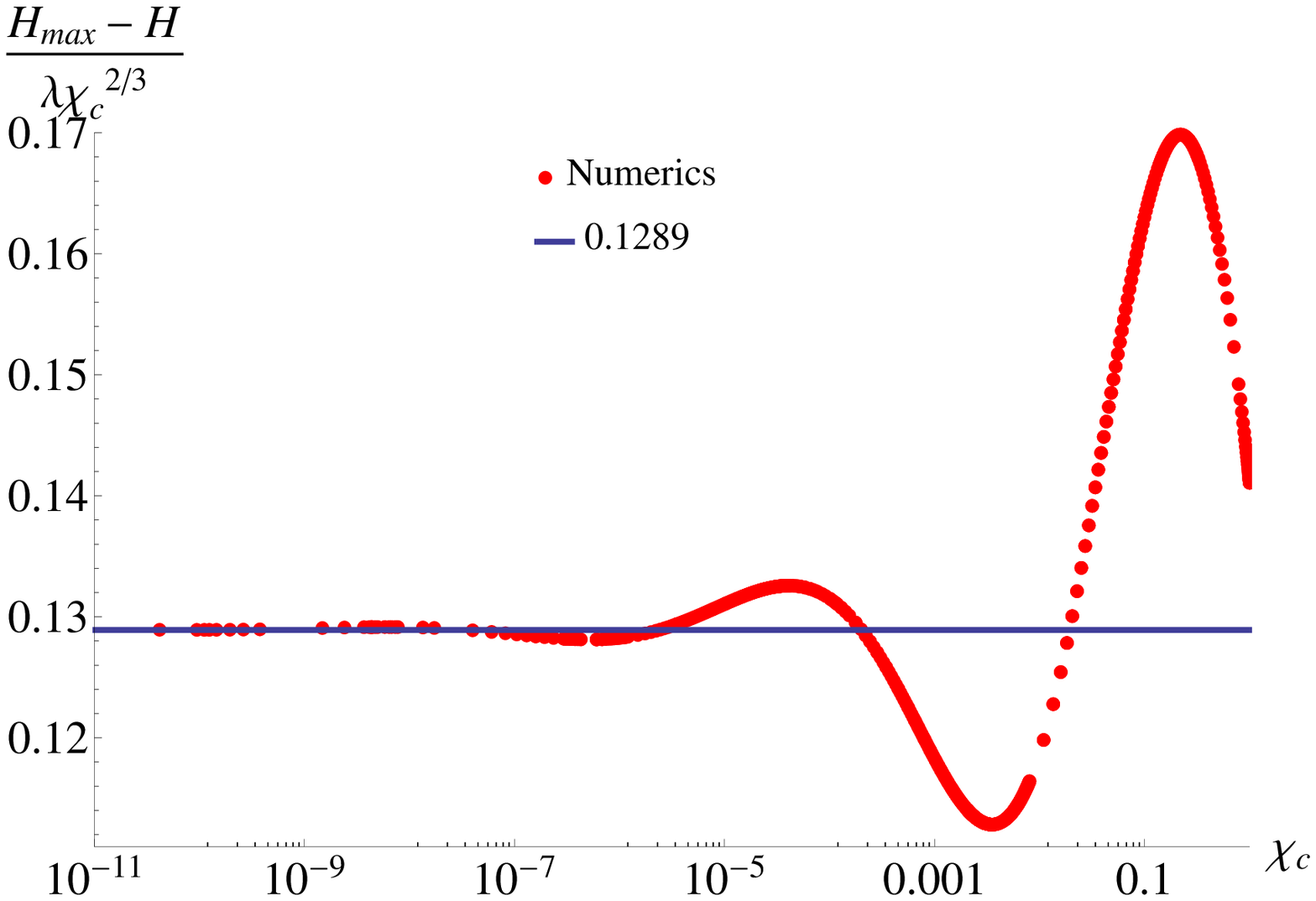}
\caption{Dependencies $\frac{c_{lim}-c}{\chi_c}$ (left panel) and  $\frac{H_{max}-H}{\lambda\chi_c^{2/3}}$ (right panel) for Stokes wave as a function of the parameter $\chi_c$. The red dots are simulation data  while the solid curves are their corresponding fits.}
\label{fig:cH2_vs_Xc}
\end{figure}

To focus on the corrections beyond the leading order scalings \e{climcscaling} and  \e{HmaxHscaling}, we plot $|c_{lim}-c|/\chi_c$  and $(H_{max}-H)/(\lambda \chi_c^{2/3})$ vs. $\chi_c$ in Fig. \ref{fig:cH2_vs_Xc}.  It seen  on left panel that  the simulation data for $(c_{lim}-c)/\chi_c$ are well fit  onto the
sin-log model

\begin{equation} \label{climmodel}
\frac{c_{lim}-c}{\chi_c}\simeq\alpha \cos [\omega_1\ln(\chi_c)+\varphi_1].
\end{equation}
 Here we used the data points from the left-most 1.5 oscillations to find  the fitting values $\alpha=0.395, \ \omega_1=0.716$
and $\varphi_1=2.01$. The points from the steepest waves are the most sensitive to the value of $c_{lim}$ on this plot. Adjusting
the value $c_{lim}$ and observing the changes in the plot while assuming that in the limit $\chi_c\rightarrow 0$ the proposed sin-log model
is valid we estimated that $c_{lim} = 1.0922850485861 \pm 5\times10^{-13}$ which is again consistent with  11 digits accuracy of  $c_{lim}^{GL}$.

One can compare Eq. \e{climmodel} with the expression%
\begin{equation} \label{Longuet-HigginsFox2}
c^2 = %(g/k)\left \{
1.1931- 1~18\epsilon^3\cos(2.143\ln\epsilon+2.22)%\right \}
\end{equation}
which was obtained by M.S. Longuet-Higgins and M.J.H. Fox in Ref.
\cite{Longuet-HigginsFoxJFM1978} by matched asymptotic expansions. Here $\epsilon:= 2^{-1/2}q$ and
$q$ is the particle speed at the wave crest in a frame of reference
moving with the phase speed $c$. To find $q$ we notice that the
complex velocity $V:=v_x-\I v_y$ is given by $V=\Pi_u/
z_u$, where $v_x$ and $v_y$ are the horizonal and vertical
velocities in physical coordinates in the rest frame and $\Pi$ is
complex potential which for Stokes wave is given by  $\Pi=c(z-w)$
(see e.g. the Appendix B of Ref.\cite{LushnikovStokesParIIJFM2016}).
It implies using the analytic solution of  Section 8 of Ref.
\cite{LushnikovStokesParIIJFM2016} that $\epsilon = 2^{-1/2}q=
2^{-1/2}|V|_{w=0}-c|= 2^{-1/2}c|( z_u-1)/
z_u-1||_{w=0}=cP\chi_c^{1/3} +O(\chi_c^{2/3}), $ where $P\sim 1$ is
the constant. Then it is seen that Eqs.  \e{climmodel} and
\e{Longuet-HigginsFox2} are consistent if we additionally notice
that $2.143\ln\epsilon\simeq0.714\ln\chi_c+const$ which is within
the accuracy of the numerical value $\omega_1=0.716$ in the
parameter  fit of Eq.  \e{climmodel}. In addition, the coefficient
$1.1931$ in right-hand side of Eq.  \e{Longuet-HigginsFox2} is the
numerical approximation of Ref. \cite{Longuet-HigginsFoxJFM1978} for
$c_{lim}^2$. Thus Fig. \ref{fig:cH2_vs_Xc} reproduces 3 oscillations
of Eq.  \e{Longuet-HigginsFox2}.

% wave example for c=1.0924 in u and q coordinates

% wave spectra for c=1.0924 in u and q coordinates

%% oscillations near extreme wave
%\begin{figure}
%\includegraphics[width = 0.45\textwidth]{steep-fig-base.pdf}
%\includegraphics[width = 0.45\textwidth]{steep-fig-new.pdf}
%\caption{Velocity oscillations near the extreme wave: (Left) velocity $c$ vs. $\frac{H}{\lambda}$; (Right) Oscillations in rescaled
%variables $\frac{c_{lim}-c}{\left( \frac{H_{max}-H}{\lambda} \right)^{3/2}}$ vs. $\ln\left[{\frac{H_{max}-H}{\lambda}}\right]$.}
%\label{fig:steep-fig}
%\end{figure}

%% plots of  Vc/(Hmax-H)^3/2  vs. Log(Vc)    (cLIM-c)/Vc vs. Log(Vc)
%\begin{figure}
%\includegraphics[width = 0.5\textwidth]{H_vs_LogVc.eps}
%\includegraphics[width = 0.5\textwidth]{c_vs_LogVc.eps}
%\caption{Dependencies $\frac{c_{lim}-c}{v_c}$ vs. $\ln(v_c)$ and  $\frac{v_c}{\left( \frac{H_{max}-H}{\lambda} \right)^{3/2}}$ vs. $\ln(v_c)$.}
%\label{fig:H_and_c_vs_LogVc}
%\end{figure}

% |c_{lim}-c| and (H_{max}-H) vs. X_c

% |c_{lim}-c|/X_c  and  (H_{max}-H)/X_c^2/3 vs. X_c

% ((H_{max}-H)/X_c^2/3 - A)/X_c^1/3  vs.  X_c           and  (c_{lim}-c)/(H_{max}-H)^3/2   vs.   Ln(H_{max}-H)
\begin{figure}
\includegraphics[width = 0.495\textwidth]{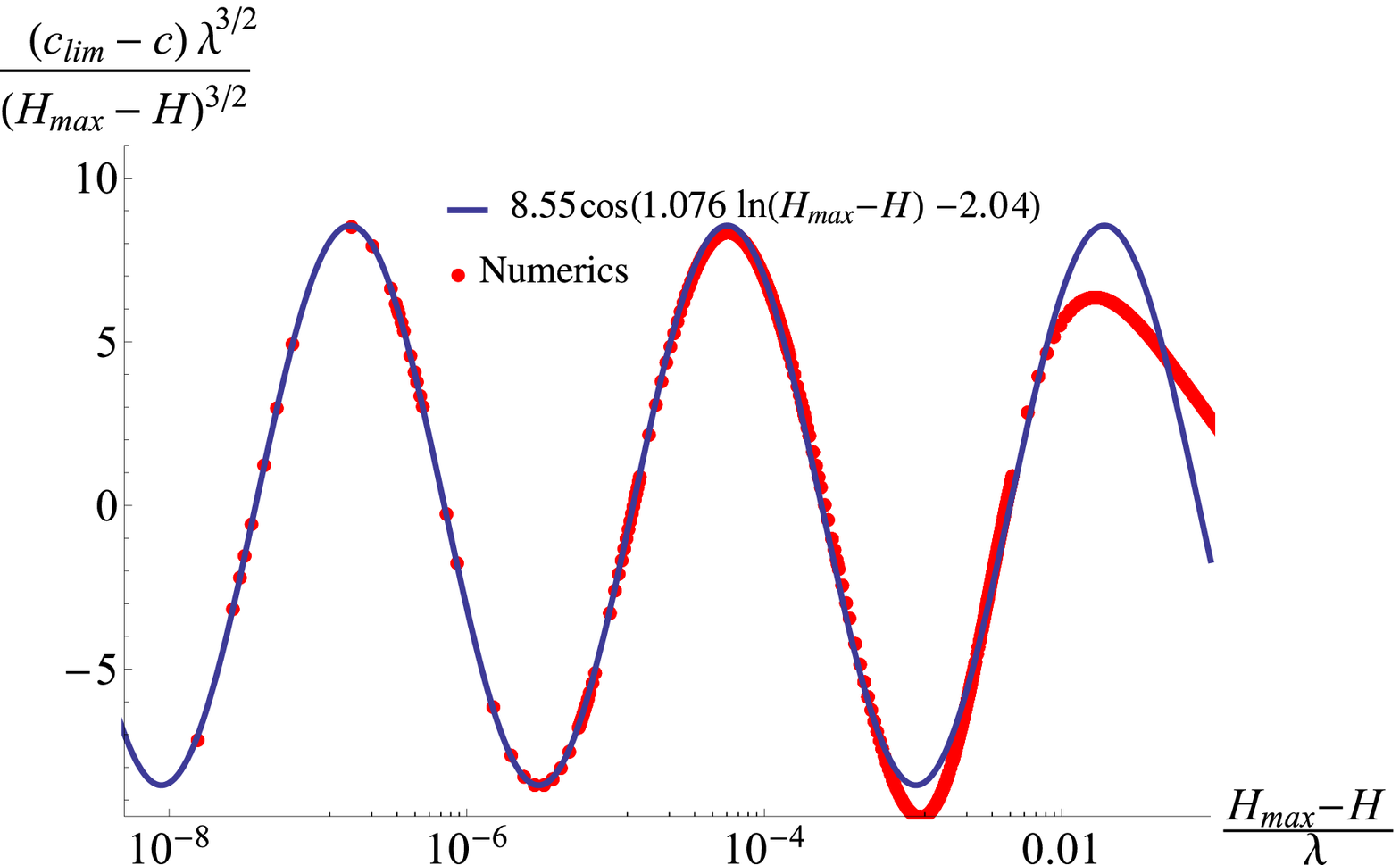}
\includegraphics[width = 0.495\textwidth]{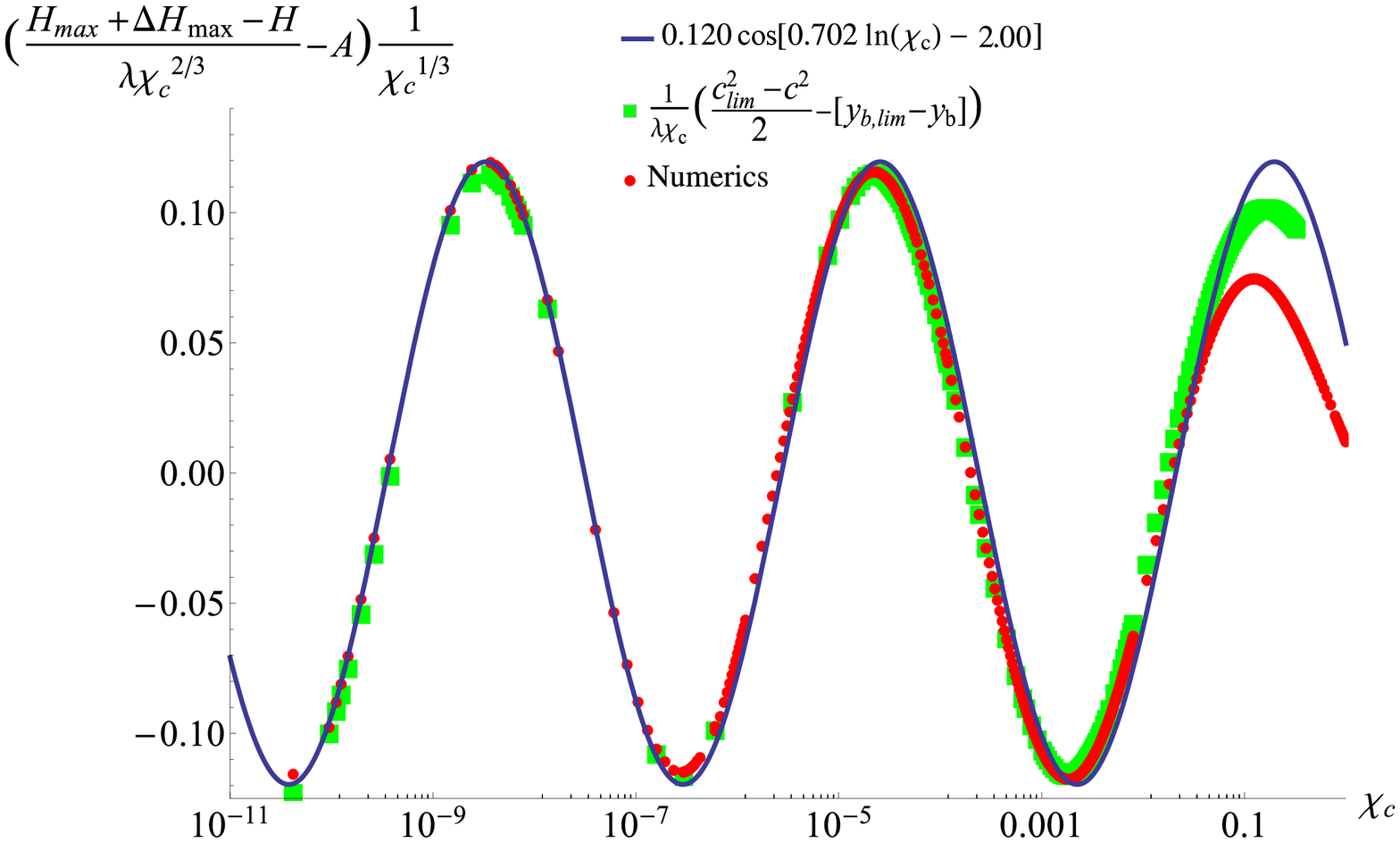}
\caption{Dependencies $\frac{(c_{lim}-c)\lambda}{(H_{max}-H)^{3/2}}$ vs. $(H_{max}-H)/\lambda$ (left panel) and $\left (\frac{H_{max}-H}{\lambda\chi_c^{2/3}}-A\right )\frac{1}{\chi_c^{1/3}}$ vs. $\chi_c$ (right panel) with $A=0.1289$.  The red dots are data points obtained from Stokes wave simulations while the solid curves are their corresponding fits.}
\label{fig:cH3_vs_Xc}
\end{figure}

The definition  of the Stokes wave height  results in $H=y(0)-y_b$, where $y_b:=y(\pm \pi)$.  Then the analytical Stokes wave solutuon of  Section 8 of Ref. \cite{LushnikovStokesParIIJFM2016} implies that the next correction       beyond the leading order model \e{HmaxAmodel}
      is given by %
\begin{equation} \label{HmaxAmodel2}
H^{GL}_{max}+\Delta H_{max}-H-\lambda A \chi_c^{2/3}=\frac{\left (c_{lim}^{GL}\right )^2-c^2}{2}-(y_{b,lim}-y_b),
\end{equation}
where $y_{b,lim}$ is the value of $y_b$ for the limiting Stokes wave (we approximate it by the most extreme wave in our simulations).
 To check the accuracy of Eq. \e{HmaxAmodel2} we divided it by  $\lambda \chi_c$ and compared the right-hand side with the left-hand side on right panel of Fig. \ref{fig:cH3_vs_Xc} showing excellent agreement. Oscillations both in $c^2(\chi_c)$ (as seen on left panel of Fig. \ref{fig:cH2_vs_Xc}) and in $y_b(\chi_c)$ are comparable in amplitude both contributing to that agreement. Inspired by the model \e{climmodel}, we fit the data $H(\chi_c)$ to the following model $H^{GL}_{max}+\Delta H_{max}-H(\chi_c)  - A \chi_c^{2/3} \simeq B \chi_c \cos[\omega_2 \ln(\chi_c) +\varphi_2]$, where  $ \Delta H_{max}, A,B,\omega_2,\varphi_2$ are unknown constants. Using the data only from the  large oscillation at the smallest $\chi_c$ on right panel of Fig.  \ref{fig:cH3_vs_Xc},  we obtained that $\Delta H_{max}/\lambda=1.3\times10^{-12}, A=0.1289, B=0.120, \omega_2=0.702, \varphi_2=-2.00$. This model fit is shown on right panel of Fig. \ref{fig:cH3_vs_Xc} by the solid curve.  Notice, that $|\omega_1|\approx|\omega_2|$. We expect that if the suggested models  are correct then $|\omega_1|$ should be equal to $|\omega_2|$ as $\chi_c \rightarrow 0$.

It is also instructive to relate the physical variables  $H_{max}-H $ and $c_{lim}-c$ directly bypassing the use of $\chi_c$. Via Eq. \e{HmaxAmodel} we approximate $\chi_c$ at the leading order through $H^{GL}_{max}-H.$ We plug in that approximation into Eq. \e{climmodel} to obtain at the leading order that
\begin{align} \label{hclim}
&(c_{lim}-c)\lambda^{3/2}/(H_{max}-H)^{3/2} \nonumber \\&\simeq
\alpha A^{-3/2} \cos\left [(3\omega_1/2)\ln[(
H^{GL}_{max}-H)/\lambda]-(3\omega_1/2)\ln{A}+\varphi_1\right ].
\end{align}

Left  panel of Fig. \ref{fig:cH3_vs_Xc} shows
$(c_{lim}-c)\lambda^{3/2}/(H_{max}^{GL}-H)^{3/2}$ vs.
$(H_{max}^{GL}-H)/\lambda$ together with the model
$(c_{lim}-c)\lambda^{3/2}/(H_{max}-H)^{3/2} \simeq D
\cos[\omega_3\ln[( H^{GL}_{max}-H)/\lambda]+\varphi_3].$  The
fitting constants of the model are $D=8.55, \ \omega_3=1.076$ and $\varphi_3=-2.04$
being consistent with the leading order expression \e{hclim}.

\section{Generalization  of the conformal map to resolve multiple singularities }
\label{sec:Generalizedconformalmap}

Assume that we aim to  approximate a general  $2\pi$ periodic  function $f(u),  \ u\in \mathbb{R}$   which has multiple complex singularities in its analytical continuation into the complex plane $w=u+\I v.$ Example of such function is  higher order progressive waves, which have more than one different peaks per $2\pi$ spatial period \cite{ChenSaffmanStudApplMath1980}. We would like to efficiently approximate $f(u)$ thought  the Fourier series in the new variable $q(u).$ A generalization of the conformal transformation \e{qnewdef}  to take into account these multiple complex singularities of the  function $f(u)$  is given by
\begin{align}\label{eqn:qucomposite}
q(u) = \sum\limits_{j=1}^N 2\beta_ j \arctan{\left[ \frac{ 1}{L_j}\left( \tan{\frac{u}{2}} -\tan{\frac{u_{j}}{2}} \right) \right]},
\end{align}
where $L_j>0$, $\beta_j>0$ and $-\pi< u_j<\pi$ are real constants. A condition
$\sum\limits_{j=1}^N\beta_j=1$ ensures that $-\pi\le q\le \pi.$ %Here we %defined  $U_{f,j}\equiv\tan{\frac{u_{f,j}}{2}} $.
The Jacobian of  Eq. \e{eqn:qucomposite} given by %
\begin{equation} \label{qujacobiancomposite1}
q_u=\sum\limits_{j=1}^N \frac{2\beta
_j}{L_j\cos^2{\frac{u}{2}} \left (1+\frac{\left
[\tan{\frac{u}{2}}-\tan{\frac{u_{j}}{2}}\right ]^2}{L_j^2}\right )}
\end{equation}
 is positive-definite. It ensures that Eq. \e{eqn:qucomposite} is one-to-one map between $-\pi\le q< \pi$ and $-\pi\le u<\pi$.

Consider an analytical continuation of  $ f(u)$ into the complex
plane $w=u+\I v$  and choose   $N$ complex singularities of $f(w)$
which are closest to the segment   $-\pi\le u<\pi$ of the real line
$w=u$. Then $u_{j}$ can be chosen either close or equal to the
projections  of the positions of complex singularities of $f(w)$
into the real line $w=u.$  Similar to the case $N=1$ described in
previous sections, the constants $L_j$ can be chosen to move the
complex singularities of $f(q)$ further away from the real line of $q$
compared with locations of singularities of $f(w)$. Then  the uniform grid in
$q$   corresponds to the nonuniform grid in $u$ concentrating at
neighborhoods of each $N $  singularities of $f(w)$ thus greatly
improving the approximation of $f(u)$ by FT in $q$. The constants
$\beta_j>0$ can be either chosen equals, $\beta_j=1/N,
j=1,2,\ldots,N,$ or nonequal (to provide stronger weights for most
dangerous complex singularities).
 Note that $u_j$ can be chosen the same for several values of $j$ (but with different values of $L_j)$ thus simultaneously resolving several singularities (or several segments of the vertical branch cut)
located along the same vertical line in $w$ plane.

Another  transformation $q(u)$ to resolve multiple singularities is given by a generalization of Eq.  \e{unewdef} as follows
\begin{align}\label{eqn:uqcomposite}
u(q) = \sum\limits_{j=1}^N 2\alpha_ j \arctan{\left[ L_j\left( \tan{\frac{q}{2}} -\tan{\frac{q_{j}}{2}} \right) \right]},
\end{align}
which is explicit expression for $u(q)$ but implicit for the inverse $q(u). $  Here $L_j>0$, $\alpha_j>0$ and $-\pi< q_j<\pi$ are real constants. A condition
$\sum\limits_{j=1}^N\alpha_j=1$ ensures that $-\pi\le q\le \pi.$ %Here we %defined  $U_{f,j}\equiv\tan{\frac{u_{f,j}}{2}}.
Similar to Eq.  \e{eqn:qucomposite},  Eq. \e{eqn:uqcomposite} has a strictly positive Jacobian which ensures one-to-one map between $-\pi\le q< \pi$ and $-\pi\le u<\pi$.

\section{Conclusion}
\label{sec:Conclusion}

In conclusion, we found the new transformation  \e{qnewdef} which allows to move the lowest complex singularity $w=\I v_c$ of the function $f(u)$ away from the real line thus  greatly improving the efficiency of FT of that function in the new variable $q$. Number of Fourier modes needed to reach the same precision of approximation of $f$ in $q$ variable is reduced by the factor $\sim v_c ^{1/2}$ for $v_c\ll 1$ compare with FT in $u$. We showed that the new transformation  \e{qnewdef} is consistent with the dynamics of two dimensional Euler equation with free surface.
 We demonstrated the effciency of Eq.   \e{qnewdef} for simulations of Stokes wave by improving the numerical performance in many orders of magnitude. It allows to reveal the details of the oscillatory behaviour of the parameters of Stokes wave  as it approaches the limiting wave. We suggested the generalizations   \e{eqn:qucomposite} and   \e{eqn:uqcomposite} of Eq. \e{qnewdef} to resolve multiple singularities.

The work of D.S. and P.L. was partially
supported by the National Science Foundation Grant DMS-1412140.

%\funding{Insert funding text here.}

%\bibliographystyle{report}
%\bibliography{surfacewaves,lushnikov,Biblio,biblionls}

\end{document}